\documentclass[pdftex,twocolumn,epjc3]{svjour3}          

\RequirePackage[T1]{fontenc}

\smartqed  

\RequirePackage{graphicx}
\RequirePackage{mathptmx}      
\RequirePackage{flushend}
\RequirePackage[numbers,sort&compress]{natbib}
\RequirePackage[colorlinks,citecolor=blue,urlcolor=blue,linkcolor=blue]{hyperref}

\usepackage{amssymb}
\usepackage{amsmath}

\journalname{Eur. Phys. J. C}

\pdfoutput=1 

\usepackage[T1]{fontenc} 

\usepackage{graphicx}

\newcommand*{\addheight}[2][.5ex]{%
  \raisebox{0pt}[\dimexpr\height+(#1)\relax]{#2}%
}

\usepackage{dcolumn}
\usepackage{bm}

\newcommand{\MSBar}{\overline{MS}}
\newcommand{\MSbar}{$\MSBar$ }

\usepackage{booktabs}

\usepackage{rotating}
\usepackage{enumitem}

\usepackage{listings}

\usepackage{colortbl}

\usepackage{siunitx}
\usepackage[dvipsnames]{xcolor}
\usepackage{booktabs,colortbl, array}
\usepackage{pgfplots,pgfplotstable,filecontents}

\definecolor{darkgreen}{rgb}{0,0.6,0}

\graphicspath{ {Figures/} {Figures_old/}}
\usepackage{pgfplots,pgfplotstable,filecontents}

\definecolor{rulecolor}{RGB}{0,71,171}
\definecolor{tableheadcolor}{gray}{0.92}
\newcommand{\topline}{ %
        \arrayrulecolor{rulecolor}\specialrule{0.1em}{\abovetopsep}{0pt}%
        \arrayrulecolor{tableheadcolor}\specialrule{\belowrulesep}{0pt}{0pt}%
        \arrayrulecolor{rulecolor}}
\newcommand{\midtopline}{ %
        \arrayrulecolor{tableheadcolor}\specialrule{\aboverulesep}{0pt}{0pt}%
        \arrayrulecolor{rulecolor}\specialrule{\lightrulewidth}{0pt}{0pt}%
        \arrayrulecolor{white}\specialrule{\belowrulesep}{0pt}{0pt}%
        \arrayrulecolor{rulecolor}}
\newcommand{\bottomline}{ %
        \arrayrulecolor{white}\specialrule{\aboverulesep}{0pt}{0pt}%
        \arrayrulecolor{rulecolor} %
        \specialrule{\heavyrulewidth}{0pt}{\belowbottomsep}}%

\pgfplotstableset{normal/.style ={%
        header=true,
        string type,
        font=\addfontfeature{Numbers={Monospaced}}\small,
        column type=l,
        every odd row/.style={
            before row=
        },
        every head row/.style={
            before row={\topline\rowcolor{tableheadcolor}},
            after row={\midtopline}
        },
        every last row/.style={
            after row=\bottomline
        },
        col sep=&,
        row sep=\\
    }
}

\newcommand{\tsil}{\textsf{TSIL} }
\newcommand{\tsils}{\textsf{TSIL}}
\newcommand{\tarcer}{\textsf{TARCER} }

\newcommand{\sarah}{\textsf{SARAH} }
\newcommand{\sarahs}{\textsf{SARAH}}
\newcommand{\feynarts}{\textsf{FeynArts} }

\newcommand{\feyncalc}{\textsf{FeynCalc} }

\newcommand{\flexiblesusy}{\textsf{FlexibleSUSY} }
\newcommand{\flexiblesusys}{\textsf{FlexibleSUSY}}
\newcommand{\fire}{\textsf{FIRE} }

\newcommand{\spheno}{\textsf{SPheno} }

\newcommand{\softsusy}{\textsf{SOFTSUSY} }

\newcommand{\looptools}{\textsf{LoopTools} }

\usepackage{slashed}
\newcommand{\mychi}{\raisebox{0pt}[1ex][1ex]{$\chi$}}

\def\Mp{M_{\mathrm{pole}}}

\begin{document}

\title{Pitfalls of iterative pole mass calculation in electroweak multiplets}

\author{James McKay\thanksref{e1,addr1},
        Pat Scott\thanksref{e2,addr1} 
                \and
        Peter Athron  \thanksref{e3,addr2,addr3}
}

\thankstext{e1}{j.mckay14@imperial.ac.uk}
\thankstext{e2}{p.scott@imperial.ac.uk}
\thankstext{e3}{peter.athron@coepp.org.au}

\institute{ Department of Physics, Imperial College London, Blackett Laboratory, Prince Consort Road, London SW7 2AZ, UK\label{addr1}
          \and
           School of Physics and Astronomy, Monash University, Melbourne, VIC 3800, Australia\label{addr2}
          \and
          Australian Research Council Centre of Excellence for Particle Physics at the Tera-scale\label{addr3}
}

\date{Received: date / Accepted: date}

\maketitle

\begin{abstract}
The radiatively-induced mass splitting between components of an electroweak multiplet is typically of order 100\,MeV. This is sufficient to endow the charged components with macroscopically-observable lifetimes, and ensure an electrically-neutral dark matter particle.  We show that a commonly used iterative procedure to compute radiatively-corrected pole masses can lead to very different mass splittings than a non-iterative calculation at the same loop order.  By estimating the uncertainties of the two one-loop results, we show that the iterative procedure is significantly more sensitive to the choice of renormalisation scale and gauge parameter than the non-iterative method.  This can cause the lifetime of the charged component to vary by up to 12 orders of magnitude if iteration is employed.  We show that individual pole masses exhibit similar scale-dependence regardless of the procedure, but that the leading scale-dependent terms cancel when computing the mass splitting if and only if the non-iterative procedure is employed. We show that this behaviour persists at two-loop order: the precision of the mass splitting improves in the non-iterative approach, but our results suggest that higher-order corrections do not reduce the uncertainty in the iterative calculation enough to resolve the problem at two-loop order.  We conclude that the iterative procedure should not be used for computing pole masses in situations where electroweak mass splittings are phenomenologically relevant.
\end{abstract}

\section{Introduction}

A fermionic multiplet coupled to the standard model (SM) via the electroweak gauge sector is a viable dark matter candidate.  However, the multiplet contains both charged and neutral components.  To explain the cosmological relic abundance of dark matter, it is therefore essential that the lifetime of the charged component be significantly shorter than the age of the universe.  This is achieved when the charged component is slightly heavier than the neutral component.  Fortunately, such a mass difference is automatically induced by radiative corrections.

Electroweak multiplet dark matter appears in several theories beyond the SM.  For instance, the wino in $R$-parity conserving supersymmetry can be a stable dark matter candidate.  If the rest of the supersymmetric spectrum is sufficiently massive to be decoupled, then a pure wino-like neutralino with a mass of $\sim$3\,TeV would give the correct relic abundance \cite{Hisano2007,Hryczuk2011}.  In this scenario, the pure wino-like chargino becomes $\sim$170\,MeV heavier than the neutralino due to radiative corrections.  This model and the radiatively-induced mass splitting have been studied extensively, including calculation of radiative corrections to the mass splitting at two-loop order \cite{Cheng1999,Feng1999,Ibe2013}.

Minimal dark matter (MDM; \cite{Cirelli2006,Cirelli2009}) is another class of models with dark matter in an electroweak multiplet.  In general, MDM refers to the assignment of a minimalistic set of quantum numbers and charges that an additional electroweak multiplet may have under the SM gauge groups.  Although most models in this class are all but ruled out \cite{Cai2015}, the fermionic quintuplet with zero hypercharge is still a viable theory.  Indeed, this particular model offers a weakly-interacting massive particle of $\sim$9\,TeV that could explain the dark matter relic abundance \cite{Cirelli2007,Cirelli2009} and solve the electroweak vacuum stability problem \cite{Chen2012a} whilst remaining almost perturbative up to the Planck scale \cite{DiLuzio2015}.  The viability of this model also relies on a radiatively-induced mass splitting of $\sim$$170$\,MeV between the neutral and charged components, and $\sim$690\,MeV between the neutral and doubly-charged components.

Radiatively-induced mass splittings are not only relevant for fermionic multiplets.  For instance, a theory with a massive spin-one vector field consisting of a charged and neutral component will also have a mass splitting of similar magnitude to the fermionic case \cite{Belyaev2017}.  Again, this is phenomenologically essential for the neutral component of the vector field to be a viable dark matter candidate.

Although a sufficiently large mass splitting is enough to provide a stable neutral dark matter candidate, determining the exact value is of significant interest for collider experiments.  The charged component of the multiplet can travel an appreciable distance within the detector of a particle collider before decaying, which would appear as a disappearing charged track.  So far, limits have only been placed on multiplet masses $\lesssim1$\,TeV, but future collider searches will probe significantly heavier scales \cite{Ostdiek2015}.  The lifetimes of charged components in a detector are extremely sensitive to the mass splitting within the multiplet.  In the wino limit of the minimal supersymmetric SM (MSSM), two-loop contributions increase the lifetime of the charged component by 10-30\% \cite{Ibe2013}.  This is because the lifetime goes as the fifth power of the mass splitting. We discuss this further in Section \ref{sec:decays}.  Therefore, it is important that the mass splitting used in any phenomenological study is as precise as possible.

To calculate the mass splitting we must determine the physical mass of the multiplet components up to a particular order in perturbation theory.  We define the physical (or \textit{pole}) mass, and outline two equivalent methods of calculating this quantity in Section \ref{sec:pole_masses}, which differ only by partial higher order corrections beyond the formal precision of the calculations.  The first is an iterative approach for finding the pole mass, which also enters the radiative corrections being applied through the external momenta. This approach has been applied in spectrum generators produced by \sarahs/\spheno \cite{Porod:2011nf,Staub:2012pb,Staub2014}, and \flexiblesusy \textsf{1.7.4} \cite{Athron2015}, which can provide a spectrum generator in any model.  The latter has the advantage that it allows one to use simple expressions that can be used for self energies of any order, making it more straightforward to extend to higher orders.  The second approach replaces the pole mass with the running scheme-dependent mass, by performing a perturbative expansion, yielding an explicit expression for the physical mass, which is truncated at the desired order.  The latter is the only method previously used to estimate mass splittings in electroweak multiplets \cite{Cirelli2006,DelNobile2010,Ibe2013,Ostdiek2015}.  We show that both approaches give equivalent values for the pole mass, with any differences between the results from the two approaches smaller than the uncertainty on the mass itself.  However, in Section \ref{sec:mass_diff} we demonstrate that the resultant mass \textit{splittings} show a significantly larger dependence on the renormalisation scale in the iterative approach than in the non-iterative method.

The large variation in the iterative mass splitting is due to logarithmic terms dependent on the renormalisation scale, which result from a large mass hierarchy.  While the physical pole mass should in principle be independent of the renormalisation scale, at the lowest orders of perturbation theory there are order-GeV variations (for a $\sim$TeV-mass particle) with respect to the choice of renormalisation scale.  Nevertheless, a one-loop mass splitting of $\sim$$170$\,MeV is often stated without an uncertainty \cite{Cirelli2006,DelNobile2010,Ostdiek2015}.  This apparent level of precision originates from an exact cancellation of scale-dependent logarithms that occur between the mass functions in the non-iterative method.  As a result, the only scale-dependence enters through the input parameters.

We show that this cancellation does not hold when using the iterative method.  If computing the mass spectrum with a renormalisation scale set to the mass of the top quark, we find a mass splitting that differs on the order of 100\,MeV from the non-iterative result for a $\sim$TeV-mass multiplet.  However, by varying the renormalisation scale we are able to account for the large hierarchy and reconcile the computational methods, albeit with a large uncertainty on the iteratively-computed mass splitting.  We also identify the origin of this difference as a remarkable transformation of the difference of one-loop functions in the large mass limit.

In perturbation theory a typical solution to an unacceptable uncertainty at one level of precision is to move to the next order.  We show that the uncertainty in the splitting prediction from the non-iterative approach is improved at two-loop order, as one would normally expect for a quantity that is not accidentally small, and find reasonable agreement with similar calculations in the literature \cite{Yamada2010,Ibe2013}.  In a companion paper \cite{McKay2017}, we compute full two-loop self energies using the non-iterative method for a range of different electroweak multiplet models, and discuss the improvements of our non-iterative two-loop calculation over those in the literature.   In this paper, we compare the results of the iterative and non-iterative calculations.  The iterative procedure for calculating the pole mass has not previously been carried out at two loops, as it leads to
infrared divergences.  However, by using a regulator mass for the photon, one can safely employ the iterative method.  However, the iterative method also requires self energies defined off shell, which are not straightforward to obtain for some diagrams.  Here we consider only a subset of diagrams, which suffice to demonstrate and understand the problem with the iterative calculation.  We show that with this partial two-loop self-energy calculation, the mass splitting exhibits a remarkably similar behaviour to the one-loop case, especially in the large mass limit.

We compute self energies for this paper using \feyncalc \textsf{9.2.0} \cite{Mertig1991,Shtabovenko2016} and \feynarts \textsf{3.9} \cite{Hahn2001}, reducing them to basis integrals with \fire \textsf{5} \cite{Smirnov2015} (via FeynHelpers \textsf{1.0.0} \cite{SHTABOVENKO201748}) and \tarcer \textsf{2.0} \cite{Mertig1998}.  We evaluate the basis integrals using \tsil \textsf{1.41} \cite{Martin2006} and analytical forms from the literature \cite{Pierce1997}.  To compute the running of the input parameters, and to cross-check the mass calculations, we generate one-loop renormalisation group equations (RGEs) and self energies with \sarah \textsf{4.8.0} \cite{Staub:2009bi,Staub:2010jh,Staub:2012pb,Staub2014} and solve them using \flexiblesusys \textsf{1.7.4}\footnote{\flexiblesusy also uses some code pieces from \softsusy \cite{Allanach2002,Allanach:2013kza}.} \cite{Athron2015}.  We have also used \sarahs/\spheno \cite{Porod:2011nf,Staub:2012pb,Staub2014} to verify the main results.

\section{Model and parameters}\label{sec:model}

For this investigation, we use a simple electroweak triplet extension of the SM.  However, our findings apply to any other model with an equivalently-induced mass splitting, such as the wino limit of the MSSM, or models with more multiplet components.  This model consists of a fermionic $SU(2)_L$ triplet $\mychi$ with zero hypercharge, coupled to the SM via the $SU(2)_L$ gauge fields.  The \MSbar renormalised Lagrangian is
\begin{align}
\mathcal{L}=\mathcal{L}_{\text{SM}}+\frac{1}{2}\overline{\mychi}\,(i\slashed{\mathcal{D}}-\hat{M})\,\mychi,
\end{align}
where $\slashed{\mathcal{D}}$ is the $SU(2)_L$ covariant derivative, $\hat{M}$ is the degenerate tree-level \MSbar multiplet mass, and $\mathcal{L}_{\text{SM}}$ is the SM Lagrangian.  At zeroth order in perturbation theory (i.e.\ tree level), the charged and neutral components have the same mass, $\hat{M}$.

We give the full one-loop self energies in a general gauge (parameterised by $\xi$) in \ref{sec:self_energies}.  The self energies are functions of $\hat{M}$, the $\MSBar$ masses of the SM gauge bosons $\hat{m}_W$ and $\hat{m}_Z$, and the $SU(2)$ gauge coupling $g$.  The self-energy functions and the input \MSbar parameters also depend on the renormalisation scale $Q$.  We use \sarah \cite{Staub2014} to generate one-loop RGEs and threshold conditions, and \flexiblesusy \cite{Athron2015} to compute the spectrum of couplings and \MSbar running masses at the required scale.

The most relevant input parameters are the physical masses $m_W = 80.404$\,GeV and $m_Z = 91.1876$\,GeV, and the coupling $\alpha^{-1}_{EM}(m_Z) = 127.934$.  For applying threshold corrections and the renormalisation group running we also require additional low energy inputs, which we take to be $m_t = 173.34$\,GeV, and fix all other parameters to the default values used in \flexiblesusys, which are kept up to date.  These have a marginal impact on the renormalisation group evolution, so we omit the details.

We evaluate the self energies in \ref{sec:self_energies} in the Landau ($\xi=0$), Feynman-'t Hooft ($\xi=1$) and Fried-Yennie ($\xi=3$) gauges.  We have also reproduced our results in the Feynman-'t Hooft gauge using self energies computed both with \sarah (4.8.0) and by hand.  We evaluate the Passarino-Veltman functions appearing in the self energies with \tsil \cite{Martin2006}, making additional checks using \looptools \cite{Romao2006}, and when possible with the integrated analytical forms from \cite{Pierce1997}.

\section{Pole mass calculations}\label{sec:pole_masses}

In this section we outline two common methods for the computation of a physical pole mass to a fixed order in perturbation theory.  The definition of a pole mass is the complex pole of the two-point propagator, which for a fermion has a denominator given by the one-particle irreducible effective two-point function
\begin{equation}
\Gamma_2=\slashed{p}-\hat{M}+\Sigma_K(p^2)\slashed{p}+\Sigma_M(p^2) \label{eqn:propagator}
\end{equation}
where $p_{\mu}$ is the four-momentum of the particle, $\hat{M}$ is the tree-level \MSbar mass and $\slashed{p}=\gamma^{\mu}p_{\mu}$.  The self energy, $\Sigma(p^2)=\Sigma_M(p^2)+\slashed{p}\Sigma_K(p^2)$, is in general a function of the renormalisation scale and any relevant masses or couplings in the theory.  We will expand the self energy in the perturbative parameter $\alpha$ such that $\Sigma^{(n)}$ is defined as $\mathcal{O}\left(\Sigma^{(n)}\right) \equiv \alpha^n$.

\subsection{The iterative pole mass}

The pole mass is obtained by demanding $\Gamma_2=0$.  This can be achieved by setting $p^2=\Mp^2$ (and $\slashed{p}=\Mp$), and solving the resulting implicit expression for the pole mass
\begin{align}
\Mp=\text{Re}\left[\frac{\hat{M}-\Sigma_M(\Mp^2)}{1+\Sigma_K(\Mp^2)}\right] \label{eqn:M_pole_iterative},
\end{align}
iteratively until the desired convergence is reached. Equivalently, one can solve
\begin{align}
\begin{split}
\Mp=\hat{M}-\Sigma_M(\Mp^2) -\Mp\Sigma_K(\Mp^2) \label{eqn:M_pole_iterative2}.
\end{split}
\end{align}
We will refer to this definition as the \textit{iterative pole mass}.

\subsection{The explicit pole mass}

While Eqs.\ (\ref{eqn:M_pole_iterative}) and (\ref{eqn:M_pole_iterative2}) can be used for self energies of any order, the same is not true of the next method that we consider.  Instead, at each higher order, the derivative of the previous loop order is required.  The iterative method therefore requires significantly less manipulation of the self energy functions than the explicit method that we will now outline.

In practice it is not always possible to use the iterative definition of the pole mass.  In such a case, one may obtain an explicit expression for $\Mp$ by making an expansion by hand in the perturbative coupling, around the tree-level mass.  We will demonstrate how such a result is obtained, working to second order in the perturbative coupling.

\begin{figure*}
\centering
\includegraphics[width=0.4\textwidth]{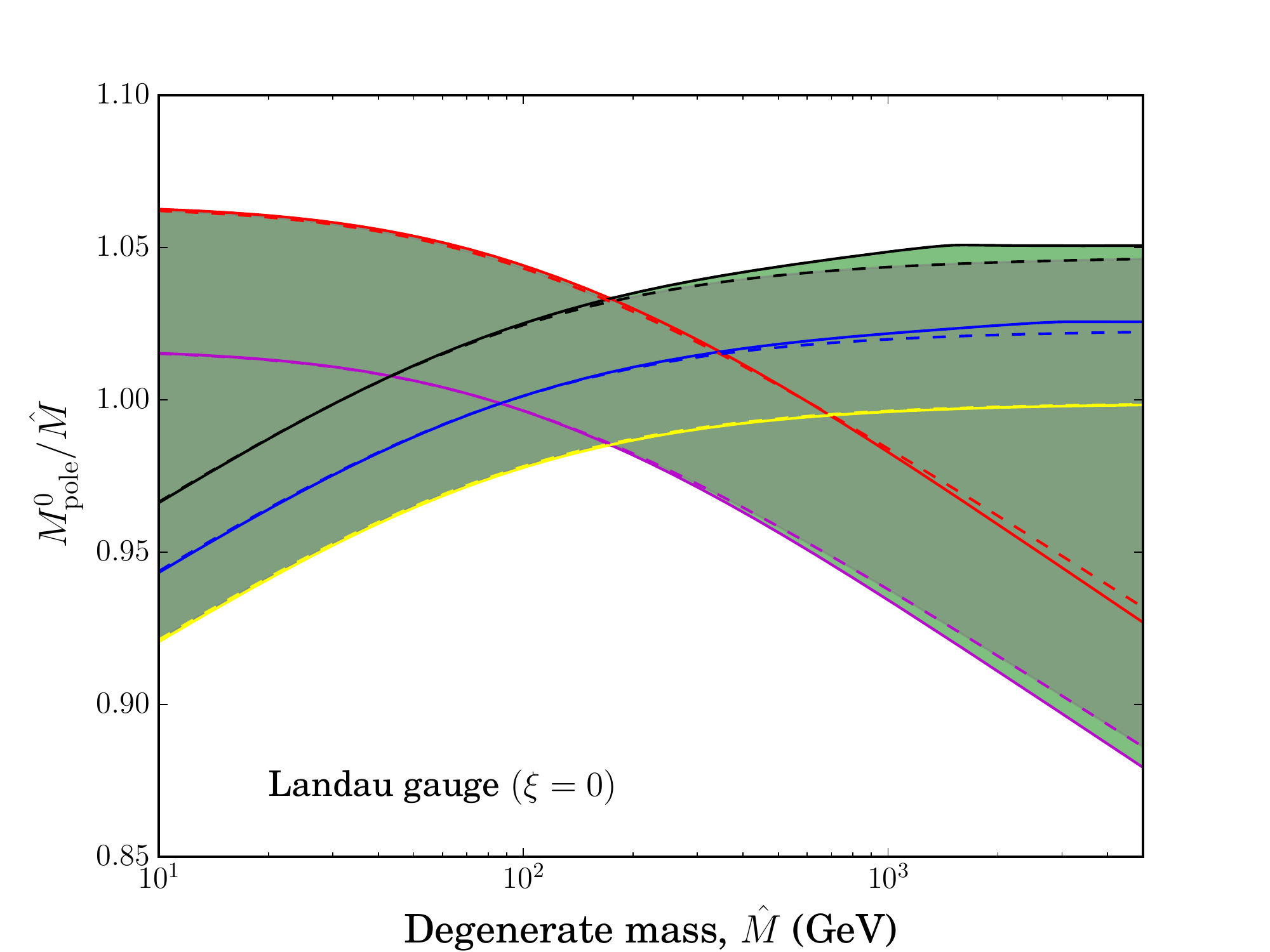}\includegraphics[width=0.4\textwidth]{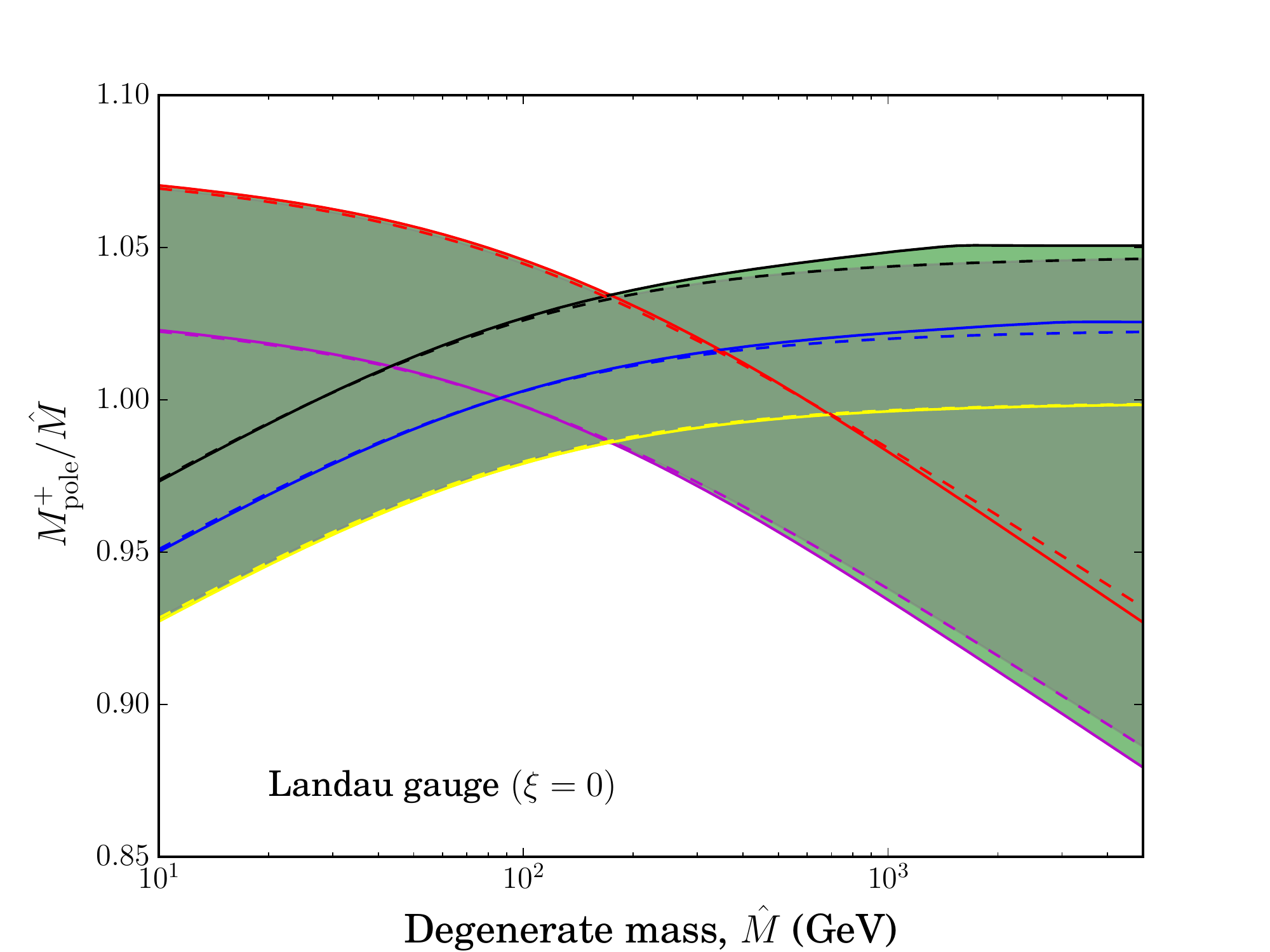}
\includegraphics[width=0.4\textwidth]{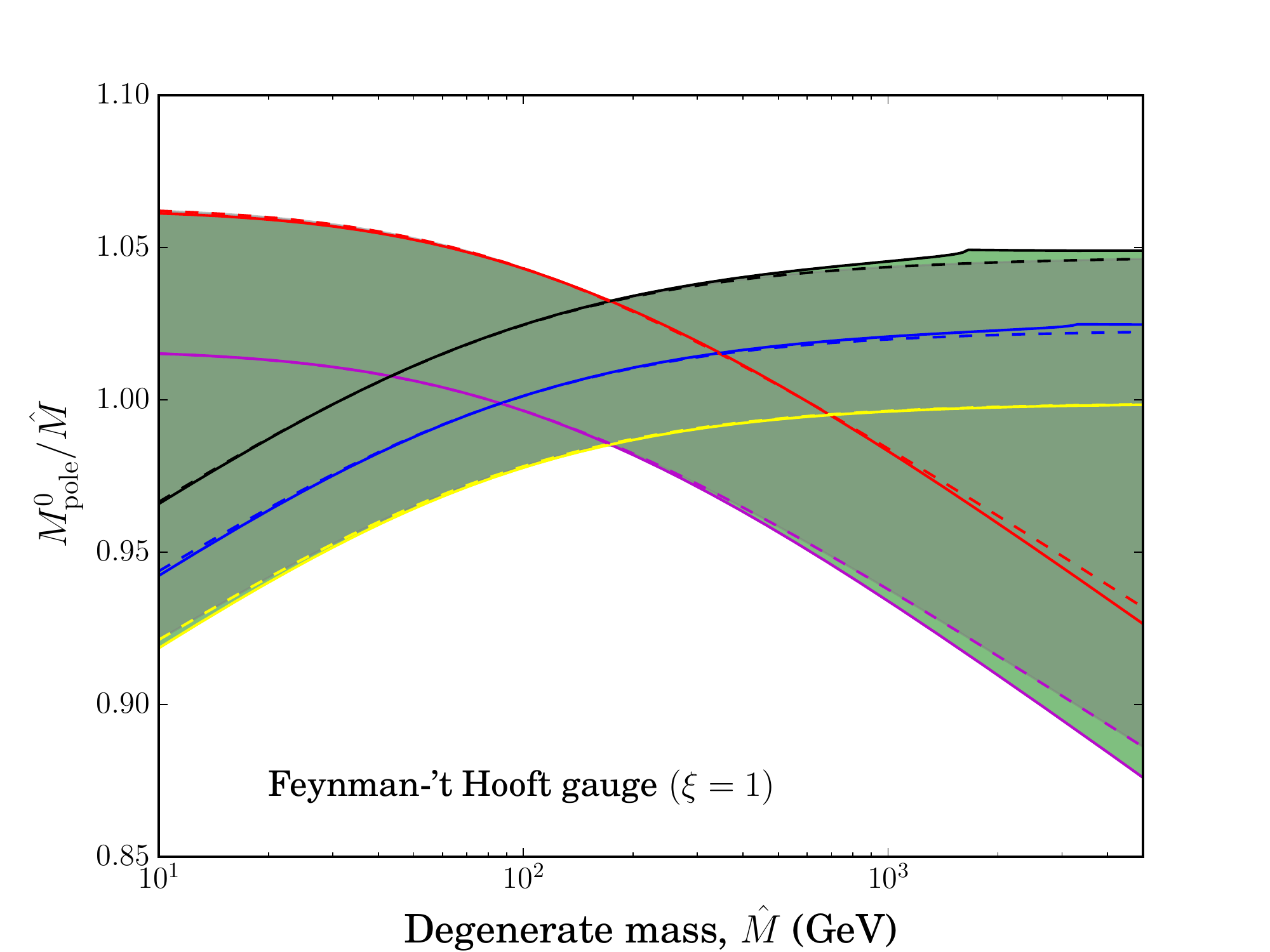}\includegraphics[width=0.4\textwidth]{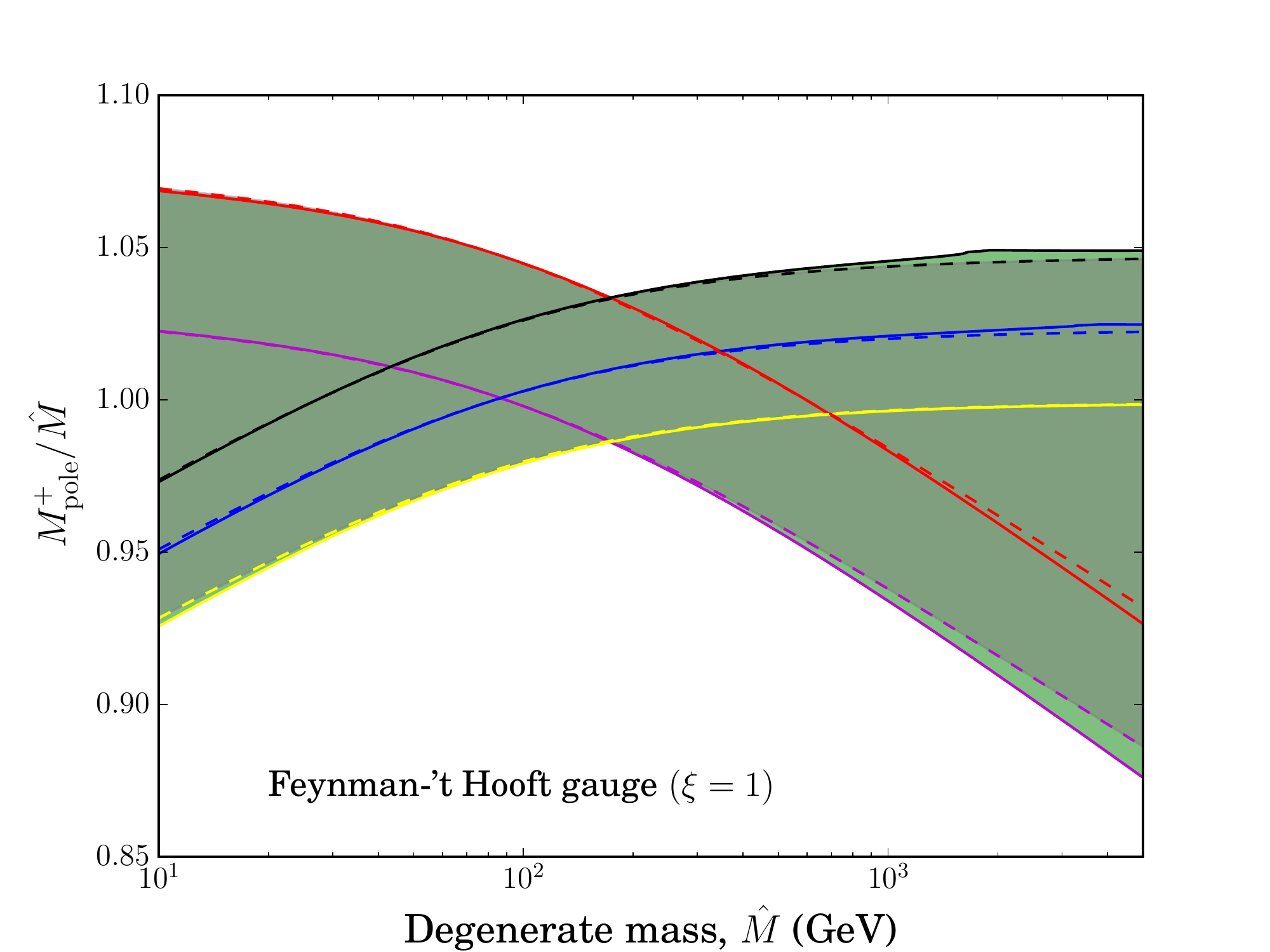}
\includegraphics[width=0.4\textwidth]{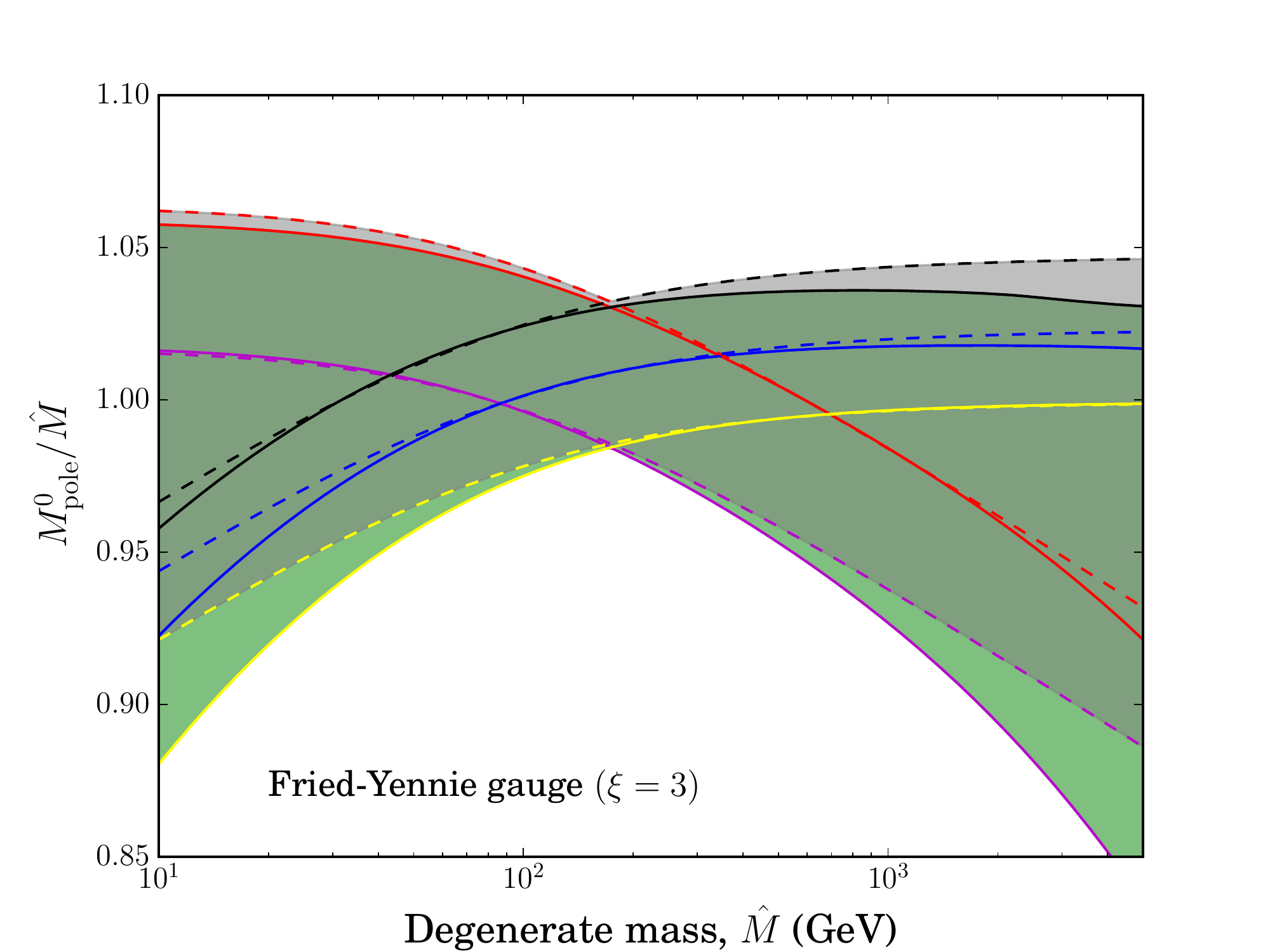}\includegraphics[width=0.4\textwidth]{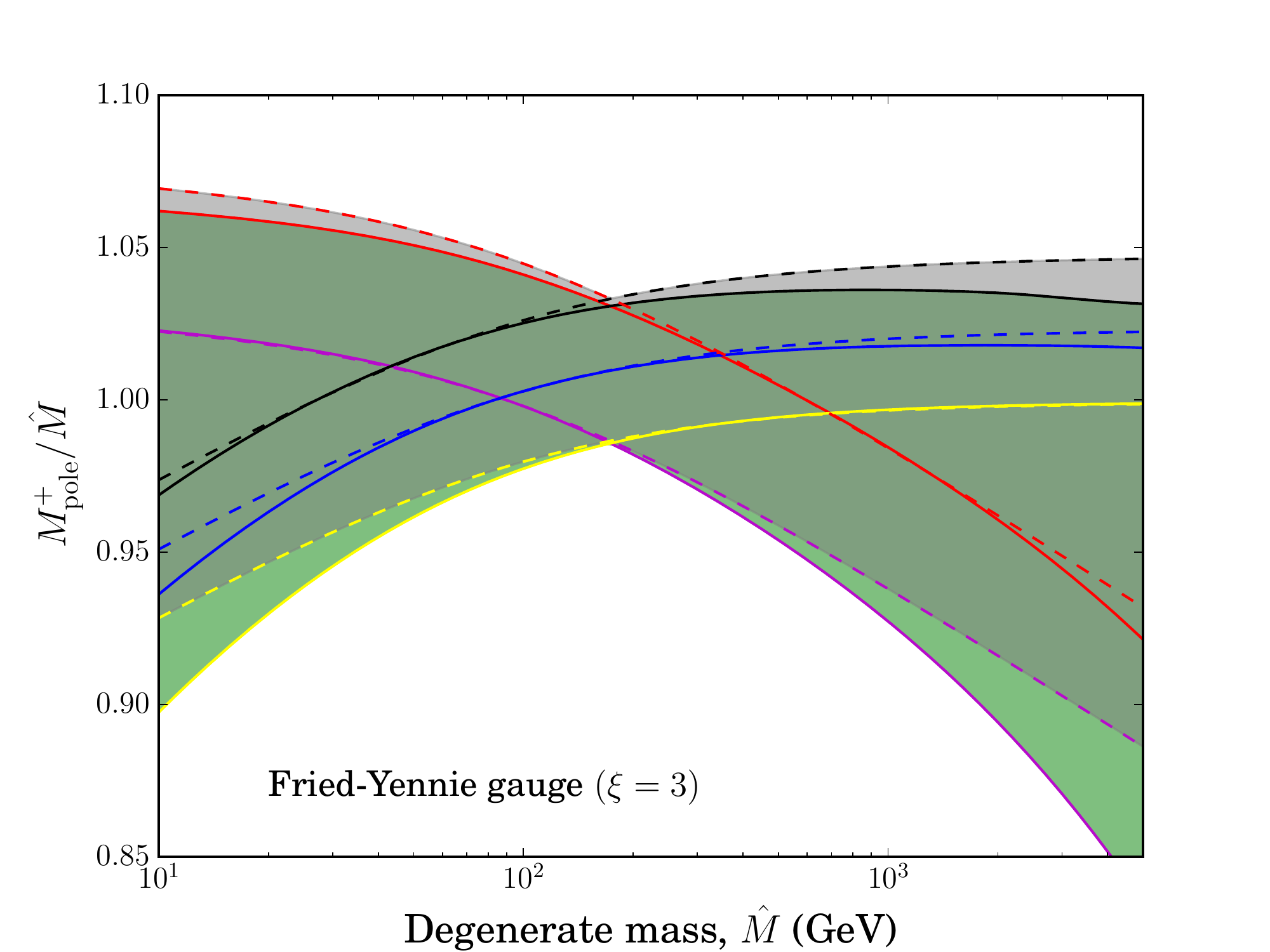}
\includegraphics[width=0.8\textwidth]{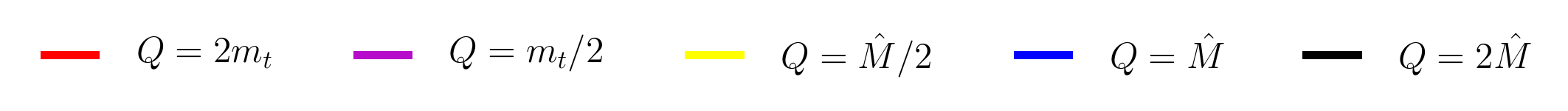}
\includegraphics[width=0.8\textwidth]{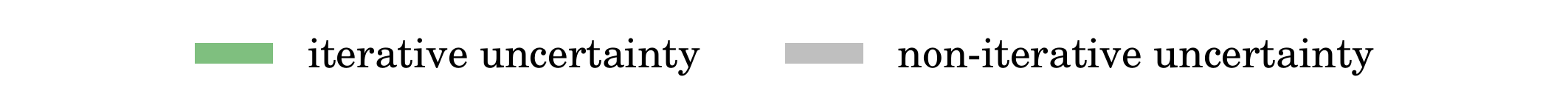}
\caption{The ratio of the one-loop pole mass to the tree-level mass for the neutral (\textit{left}) and charged (\textit{right}) components of the electroweak triplet.  The solid lines indicate values computed using the iterative method, (Eq.\ \ref{eqn:M_pole_iterative}), and dashed lines the explicit method (Eq.\ \ref{eqn:M_pole_explicit}), at fixed values of the renormalisation scale $Q$.  The shaded bands indicate the range of values obtained by varying $Q$ continuously between $\min\{\hat{M}/2,m_t/2\}$ and $\max\{2\hat{M},2m_t\}$, for each calculation method.}\label{fig:pole_masses}
\end{figure*}

Demanding that the self energies are evaluated at the tree-level mass requires the use of the Taylor expansion
\begin{align}
\begin{split}
&\left.\Sigma^{(1)}_M \right|_{p^2=\Mp^2}=\\&\left.\Sigma^{(1)}_M+2\hat{M}(\Mp-\hat{M})\dot{\Sigma}^{(1)}_M+\mathcal{O}\left(\alpha^3\right)\right|_{p^2=\hat{M}^2} \label{eqn:Taylor_expansion1},
\end{split}
\end{align}
where $\Sigma^{(n)}_X=\Sigma^{(n)}_X(p^2)$ and an over-dot, $\dot{\Sigma}$, indicates a derivative with respect to the external momentum squared. However one term on the right hand side still includes $\Mp$, so we use the relation
\begin{align}
\begin{split}
\Mp-\hat{M}&=-\Mp\Sigma^{(1)}_K(\Mp^2)-\Sigma^{(1)}_M(\Mp^2)+\mathcal{O}(\alpha^2)\\
&=-\Mp\Sigma^{(1)}_K(\hat{M})-\Sigma^{(1)}_M(\hat{M})+\mathcal{O}(\alpha^2),\label{eqn:mass_diff}
\end{split}
\end{align}
which comes directly from demanding $\Gamma_2$ be equal to zero (from Eq.\ \ref{eqn:propagator}), and the second line follows from Eq.\ (\ref{eqn:Taylor_expansion1}).  An error of order $\alpha^2$ is acceptable for this difference, as it appears in the final expression as the coefficient of $\dot{\Sigma}^{(1)}_K$, so will contribute to a total error of order $\alpha^3$.  To remove the remaining $\Mp$ on the right-hand side of Eq.\ (\ref{eqn:mass_diff}), we use the same expression within itself (effectively iterating once by hand) to obtain
\begin{align}
\begin{split}
\Mp-\hat{M}&=-\hat{M}\Sigma^{(1)}_K(\hat{M})-\Sigma^{(1)}_M(\hat{M})+\mathcal{O}(\alpha^2)\label{eqn:M0Mp}.
\end{split}
\end{align}
We then substitute this expression into Eq.\ (\ref{eqn:Taylor_expansion1}) to give
\begin{align}
\begin{split}
&\left.\Sigma^{(1)}_M\right|_{p^2=\Mp^2}=\\ &\left.\Sigma^{(1)}_M-2\hat{M}(\hat{M}\Sigma^{(1)}_K+\Sigma^{(1)}_M)\dot{\Sigma}^{(1)}_M+\mathcal{O}\left(\alpha^3\right)\right|_{p^2=\hat{M}^2}.
\end{split}
\end{align}

For the two-loop self energy functions, we can immediately take $\Sigma^{(2)}_M(\Mp^2)=\Sigma^{(2)}_M(\hat{M}^2)+\mathcal{O}\left(\alpha^3\right)$, as the derivative terms will be of higher order.  Similar relations hold for $\Sigma^{(1)}_K$ and $\Sigma^{(2)}_K$.  Finally, we can express the pole mass valid to order $\alpha^2$ as
\begin{align}
\begin{split}
&\Mp=\left[\hat{M}-\Sigma^{(1)}_M-\Sigma^{(2)}_M-\hat{M}\Sigma^{(1)}_K-\hat{M}\Sigma^{(2)}_K\right.\\
&+(\Sigma^{(1)}_M+\hat{M}\Sigma^{(1)}_K)(\Sigma^{(1)}_K+2\hat{M}\dot{\Sigma}^{(1)}_M+2\hat{M}^2\dot{\Sigma}^{(1)}_K)\\
&\left.+\mathcal{O}\left(\alpha^3\right)\right]_{p^2=\hat{M}^2}, \label{eqn:M_pole_explicit}
\end{split}
\end{align}
which is the second method of pole mass calculation that we will use. We refer to this as the \textit{explicit}, or \textit{non-iterative}, \textit{pole mass}.

Truncating Eq.\ (\ref{eqn:M_pole_explicit}) to first order in $\alpha$ gives a simple expression for the one-loop pole mass.  However, the two-loop result requires expressions for the derivatives of the one-loop functions.  In general these are not simple to obtain and implement, making the iterative approach more attractive.  On the other hand, as we will discuss in Section \ref{sec:two_loop}, it is not always possible to obtain two-loop self energies defined away from $p^2=\hat{M}^2$, making the use of Eq.\ (\ref{eqn:M_pole_explicit}) mandatory.

In Figure \ref{fig:pole_masses}, we present the one-loop pole masses for the charged, $\Mp^+$, and neutral, $\Mp^0$, components of the electroweak triplet, for three different gauge choices.  Due to the scales involved, we present the pole masses in terms of their ratios to the tree-level mass.  We show pole masses computed for $Q = m_t/2, \, 2m_t,\, \hat{M}/2,\, \hat{M}$ and $2\hat{M}$.  We obtain uncertainty bands by smoothly varying the renormalisation scale continuously between $\min\{\hat{M}/2,m_t/2\}$ and $\max\{2\hat{M},2m_t\}$.  There is a large variation in both the iterative and non-iterative pole masses as the renormalisation scale is changed.  Any discrepancy between the two methods is small, however, relative to the magnitude of this uncertainty.  There therefore appears no reason to favour one method over the other, at the level of pole masses themselves.  However, as we will show in Section \ref{sec:mass_diff}, the non-iterative pole mass produces remarkably different results when the \textit{difference} between the charged and neutral masses is considered instead.

\section{The mass splitting}\label{sec:mass_diff}

\begin{figure*}
\centering
\includegraphics[width=0.5\textwidth]{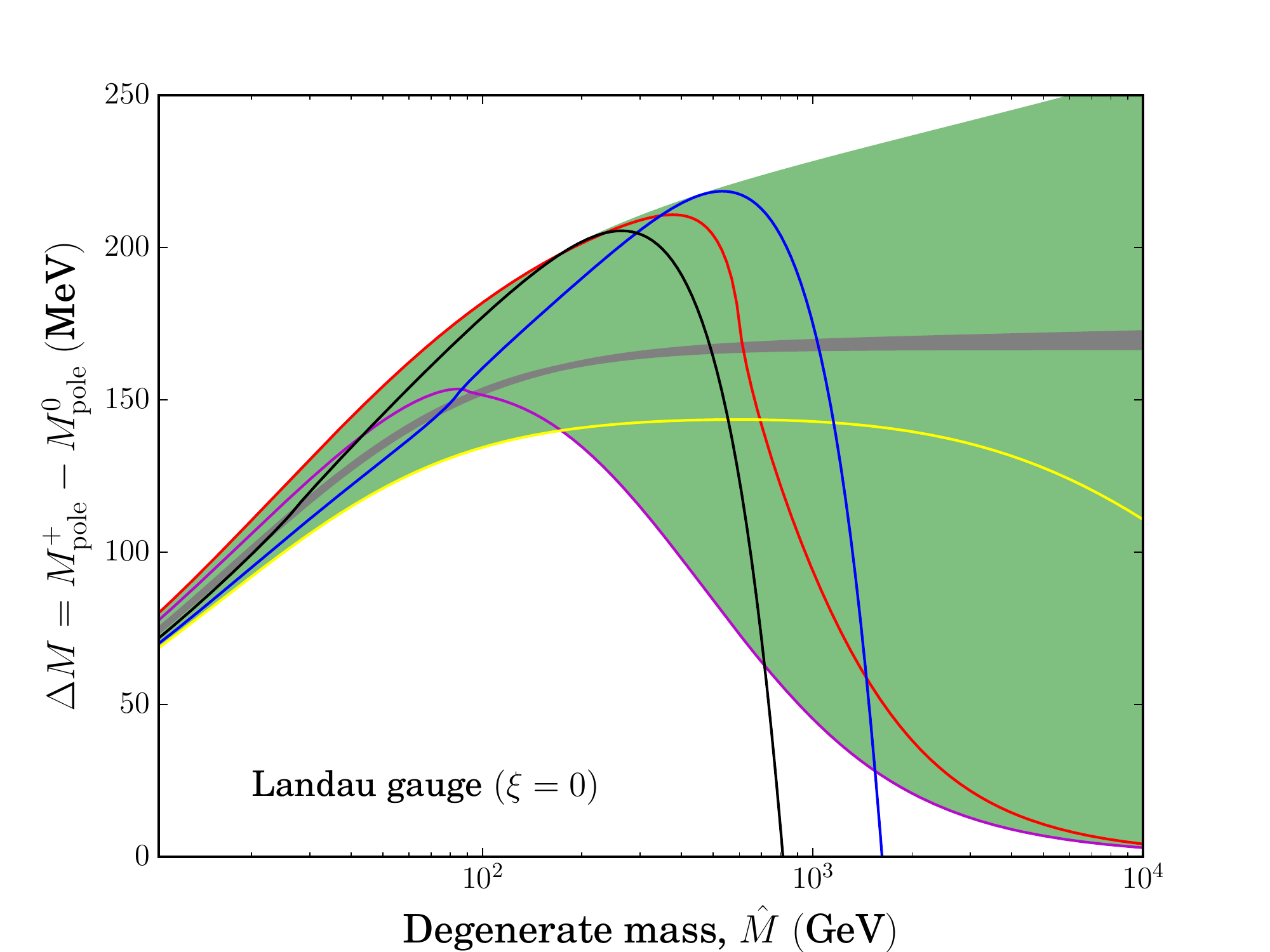}\includegraphics[width=0.5\textwidth]{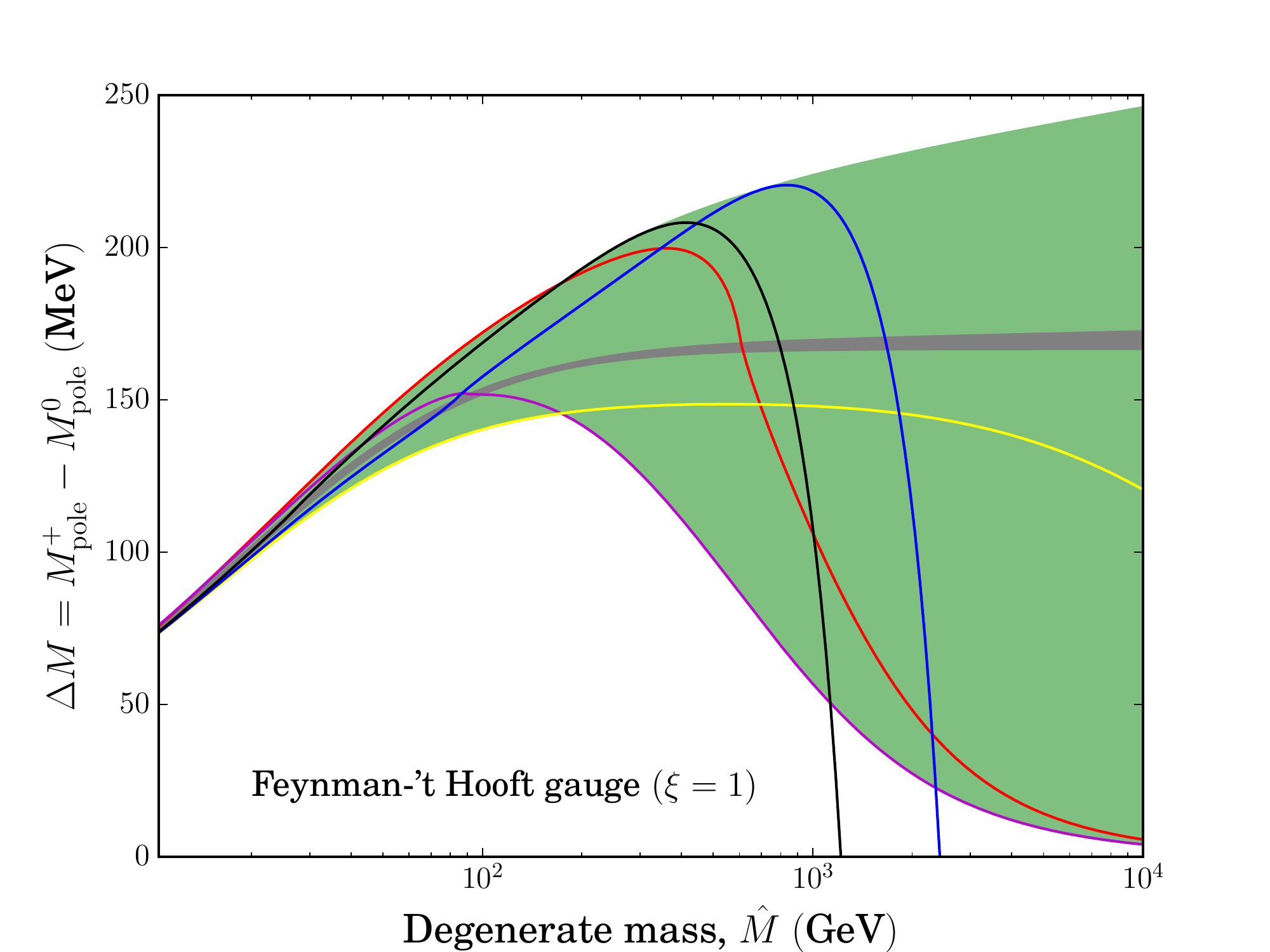}\\
\includegraphics[width=0.5\textwidth]{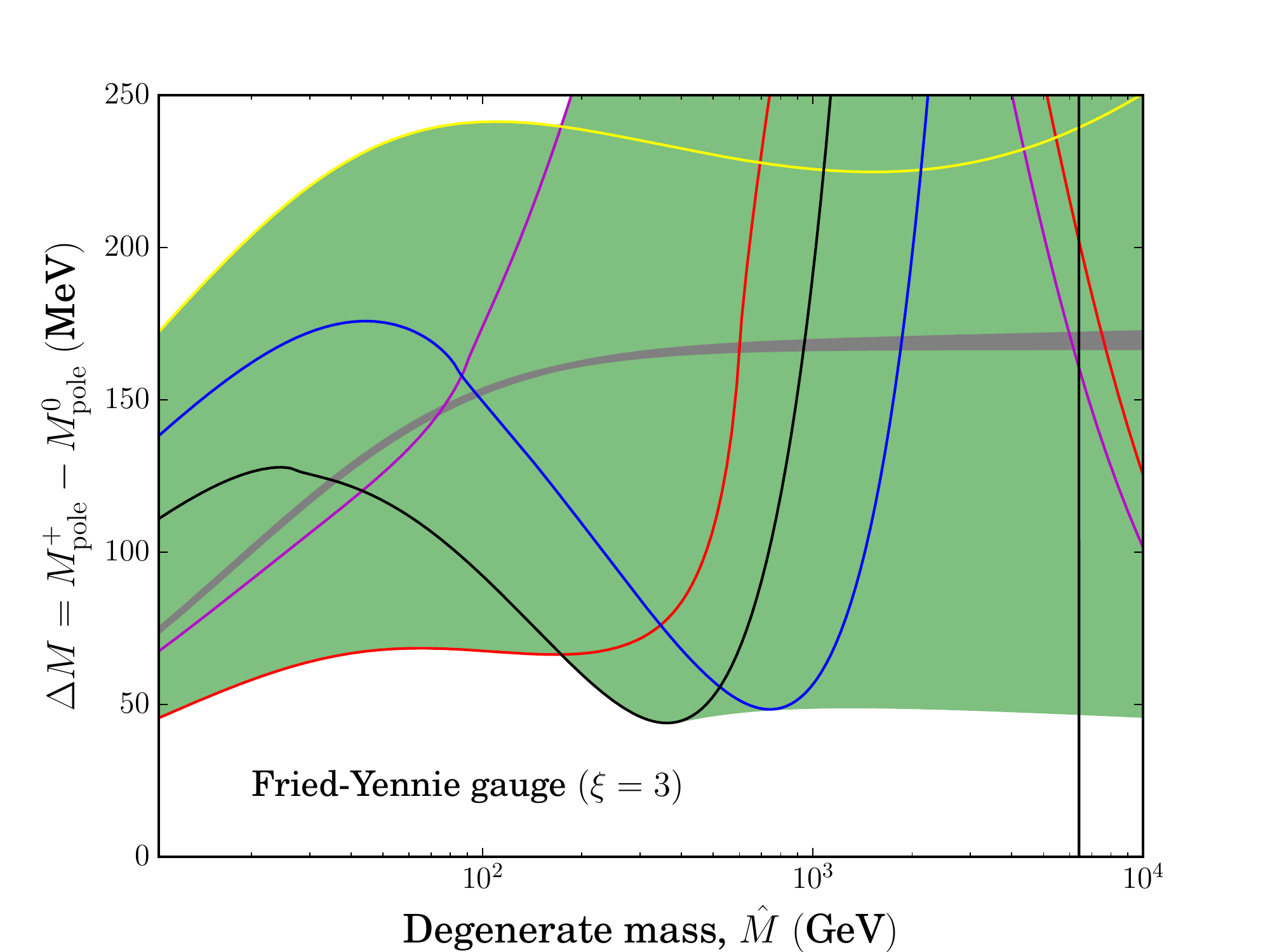}
\includegraphics[width=0.9\textwidth]{legend_1.pdf}
\includegraphics[width=0.9\textwidth]{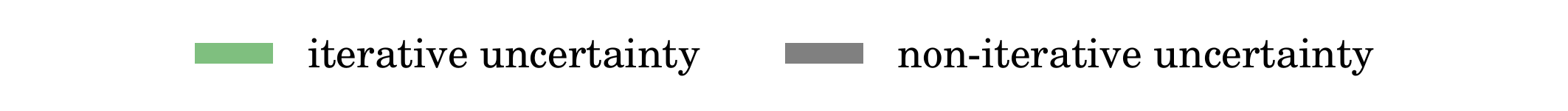}
\caption{The one-loop mass splitting $\Delta M\equiv \Mp^+-\Mp^0$ as a function of the degenerate mass $\hat{M}$.  The solid lines represent $\Delta M$ computed using the iterative method (Eq.\ \ref{eqn:M_pole_iterative}), at fixed values of the renormalisation scale $Q$.  The equivalent lines for the explicit method are entirely contained within the grey uncertainty region, so we omit them.  The shaded bands indicate the range of values obtained by varying $Q$ continuously between $\min\{\hat{M}/2,m_t/2\}$ and $\max\{2\hat{M},2m_t\}$, for each calculation method.
}\label{fig:deltam}
\end{figure*}

In Figure \ref{fig:deltam} we present the mass splitting \mbox{$\Delta M\equiv \Mp^+-\Mp^0$} as a function of the degenerate tree-level mass $\hat{M}$, again for three different choices of gauge.  We compute the iterative pole masses, and resultant mass splittings, at renormalisation scales $Q = m_t/2$, $2m_t$, $\hat{M}/2$, $\hat{M}$ and $2\hat{M}$.  For each value of $\hat{M}$, we again determine an uncertainty band by varying the renormalisation scale continuously between $\min\{\hat{M}/2,m_t/2\}$ and $\max\{2\hat{M},2m_t\}$, and taking the uncertainty to encompass the minimum and maximum mass splittings determined in each computational method.

For very large values of the renormalisation scale, the iterative mass splitting reaches a maximum at some value of $\hat{M}$, before suddenly dropping to negative values at larger $\hat{M}$. An example of this can be seen in the $Q=\hat{M}$ and $Q=2\hat{M}$ curves in Figure \ref{fig:deltam}, shown in blue and purple, respectively.  In such cases, we consider the result too unreliable, as negative values of the mass splitting are unphysical (they would lead to charged dark matter and violate the classical argument discussed in Section \ref{sec:conclusion}).  At a given value of $\hat{M}$, we therefore do not include data for values of $Q$ leading to $\Delta M < 0$ when computing the uncertainty band.  We note however that it is still very important to consider values of $Q>\hat{M}/2$ for establishing the upper bound on the iterative mass splitting, with values closer to $\hat{M}/2$ remaining close to the non-iterative result at larger and larger mass scales.

\subsection{The explicit mass splitting}

The weak dependence of the explicit mass splitting on the renormalisation scale can be understood by a symbolic calculation in the limit $\hat{M}\gg m_{Z,W}$.  The one-loop self energies are given in terms of Passarino-Veltman (PV) \cite{tHooft1979,Passarino1979} functions $B_0(p,M,m)$ and $A_0(M)$, defined in \ref{sec:self_energies}.  When using the explicit pole mass Eq.\ (\ref{eqn:M_pole_explicit}) these functions are evaluated at $M=\hat{M}$, $m=\hat{m}_{Z,W}$ and $p^2=\hat{M}^2\gg m$.  In this case the limits are given by \cite{Ibe2013}
\begin{eqnarray}
B_0(\hat{M},\hat{M},m)&=&\Delta-\log\left(\frac{\hat{M}^2}{Q^2}\right)+2-\pi \frac{m}{\hat{M}}\\&&+\mathcal{O}\left(\frac{m^2}{\hat{M}^2}\log\frac{\hat{M}^2}{m^2}\right) \nonumber \\
\frac{A_0(\hat{M})}{\hat{M}^2}&=&\log\left(\frac{\hat{M}^2}{Q^2}\right) -1 + \Delta ,\label{eqn:large_M_limit}
\end{eqnarray}
where $\Delta=2/(4-d)-\gamma+\log(4\pi)$ is cancelled by the appropriate counter-terms (see \ref{sec:self_energies}).  With the use of these limits the mass splitting becomes
\begin{align}
 \lim_{\hat{M}\gg m_{Z}}\Delta M=&\frac{g^2}{8\pi}(m_W-c_W^2 m_Z)\approx 165\ \text{MeV}, \label{eqn:simple_mass_splitting}
\end{align}
which agrees with Ref.\ \cite{Cirelli2009}.  Here $c_W=\cos(\theta)$ is the cosine of the Weinberg angle and we have taken $\hat{m}_W=m_W$ and $\hat{m}_Z=m_Z$ since threshold corrections to these masses are of next loop order.  In Eq.\ (\ref{eqn:simple_mass_splitting}), all logarithms of the form $\log(m_X/Q^2)$, where $m_X \in \{\hat{M},\hat{m}_W,\hat{m}_Z\}$, have cancelled exactly, leaving the only renormalisation dependence coming from the gauge coupling.

\subsection{The iterative mass splitting}

We find that the iterative mass splitting is highly dependent upon the chosen renormalisation scale.  Although it is not possible to write down an analytical expression analogous to Eq.\ (\ref{eqn:simple_mass_splitting}) that would be at all tractable, we can show that the limits used in Eq.\ (\ref{eqn:large_M_limit}) do not hold in the iterative case.

\begin{figure}
\centering
\includegraphics[width=0.5\textwidth]{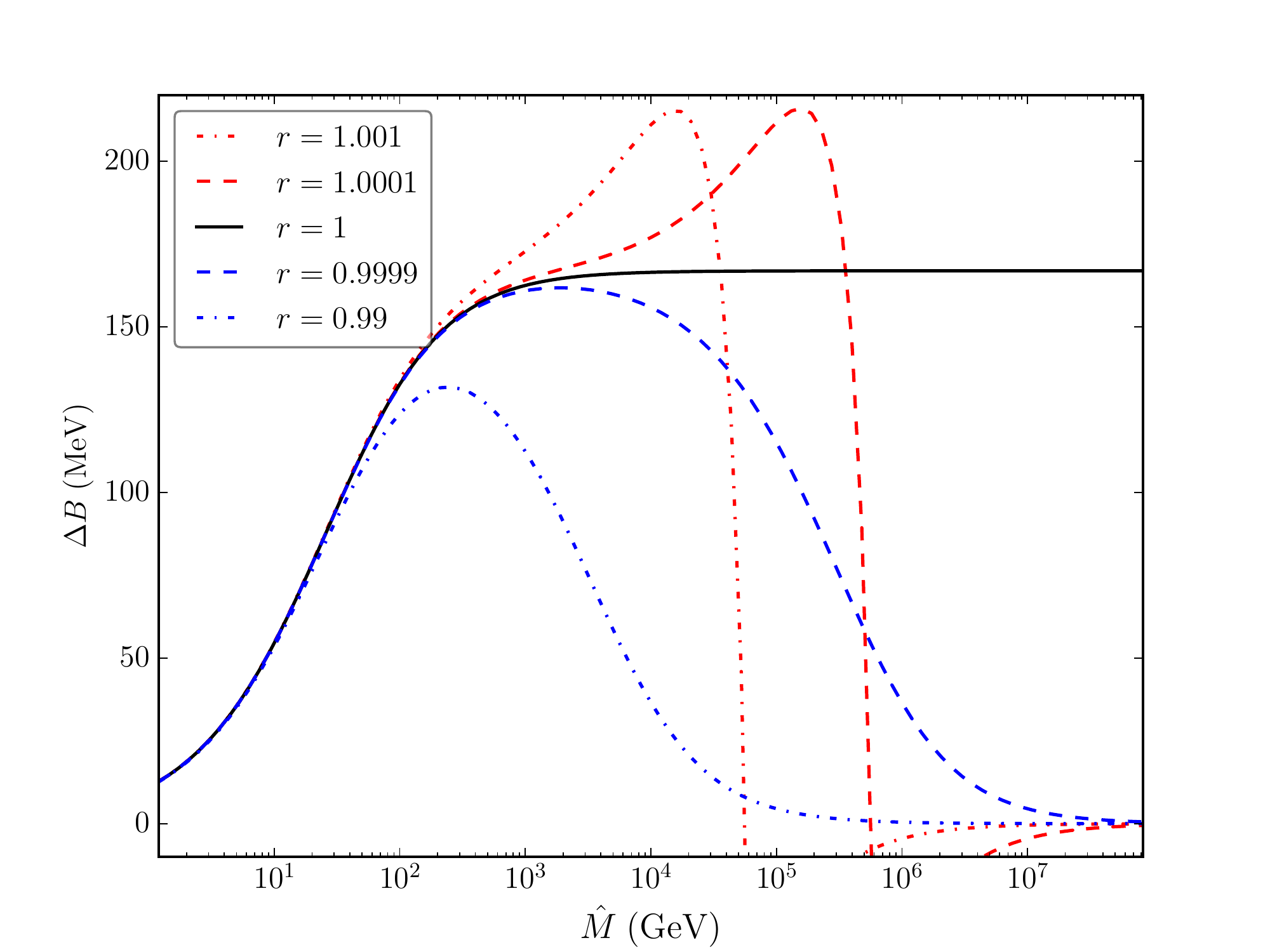}
\caption{The value of $\Delta B$ from Eq.~(\ref{eqn:DeltaB}) as a function of $\hat{M}$ for different choices of $r$.}\label{fig:DeltaB}
\end{figure}

\begin{figure}
\centering
\includegraphics[width=0.5\textwidth]{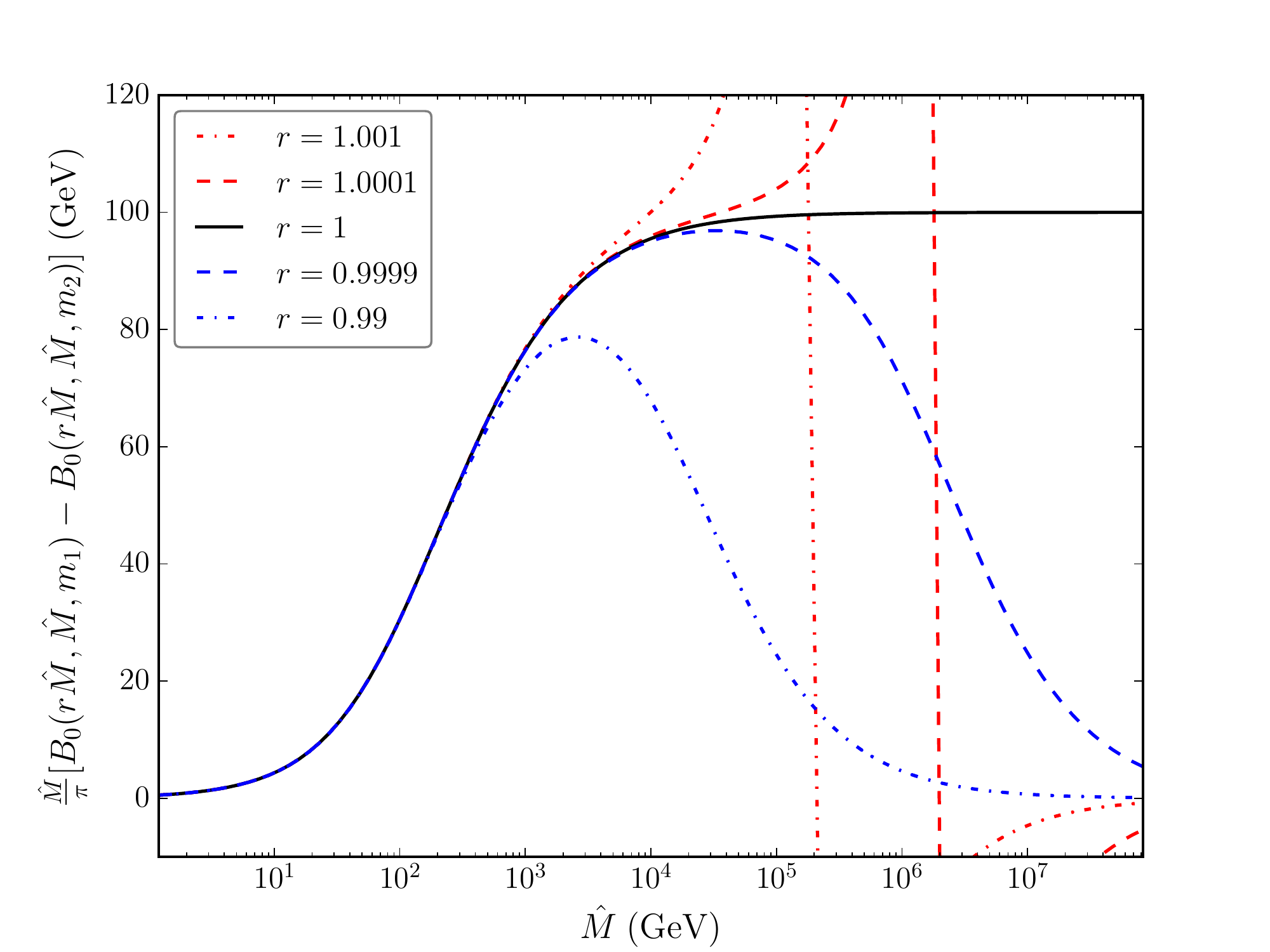}
\caption{The value of $\frac{\hat{M}}{\pi}\left[B_0(r\hat{M},\hat{M},m_1)-B_0(r\hat{M},\hat{M},m_2)\right]$ as a function of $\hat{M}$ for different choices of $r$, with $m_1 = 100\,$GeV, $m_2 = 200\,$GeV.}\label{fig:r_diff}
\end{figure}

When the iterative pole mass has converged to $\Mp$, it can be expressed as a function of the self energy evaluated at $p^2=\Mp^2$.  The self energy then becomes not only a function of $\hat{M}$, but \textit{also} an implicit function of $\Mp$.  Making the approximation $\Mp \equiv \Mp^+\approx\Mp^0$, and neglecting all terms which become small in the limit $\hat{M}\gg m_Z$ (i.e. terms of order one or more in $1/\hat{M}$ or $1/p$) we are left with the iterative mass splitting
\begin{align}
\lim_{\hat{M}\gg m_Z}\Delta M =& \frac{g}{16\pi^2}\left(  4\hat{M} - \frac{\hat{M}^2}{\Mp}-\Mp    \right)\left[ B_0(\Mp,\hat{M},m_W)\right.\nonumber\\
&\left.-s_W^2B_0(\Mp,\hat{M},0)- c_W^2 B_0(\Mp,\hat{M},m_Z) \right]\nonumber\\
= & \frac{\hat{2M}g}{16\pi^2}\frac12\left(4-\frac1r-r\right)\left[  B(r\hat{M},\hat{M},m_W)\right.\label{eqn:DeltaB}\\
&\left.-s_W^2B_0(r\hat{M},\hat{M},0)- c_W^2 B_0(r\hat{M},\hat{M},m_Z) \right]\nonumber\\
\equiv&\ \Delta B,\nonumber
\end{align}
where $r\equiv M_{\mathrm{pole}}/\hat{M}$.  For $r$ within any realistic distance of 1, the prefactor $\frac12\left(4-\frac1r-r\right)$ is close to 1, and has little impact on the splitting.  However, the fact that $r\neq1$ in the argument to the $B_0$ integral has significant implications for the evaluation of the mass splitting.  We plot $\Delta B$ in Figure \ref{fig:DeltaB} as a function of $\hat{M}$ for different values of $r$.  For $r=1$ and $\hat{M}\gg m_Z$, this expression approaches $\sim$$170\,$MeV, analogous to the result in Eq.\ (\ref{eqn:simple_mass_splitting}).  However, for $r\neq 1$ the large $\hat{M}$ behaviour of Eq.\ (\ref{eqn:DeltaB}) is significantly different.  Similar behaviour can also be seen in a simpler combination of $B_0$ functions,
\begin{align}
\frac{\hat{M}}{\pi}\left[B_0(r\hat{M},\hat{M},m_1)-B_0(r\hat{M},\hat{M},m_2)\right]\label{eqn:r_diff2}.
\end{align}
which we plot in Figure \ref{fig:r_diff}. One can immediately see similarly remarkable differences in the large $\hat{M}$ limit from small variations in $r$.

It is evident from Figure \ref{fig:DeltaB} that the limit of Eq.\ (\ref{eqn:simple_mass_splitting}) does not hold in the case that $r\neq 1$.  Instead, the difference appears to approach the $\sim$$170\,$MeV limit with increasing $\hat{M}$, until a critical point is reached.  If $r<1$ then a turn-over occurs, beyond which the curve asymptotically approaches zero.  If $r>1$, then there is a rapid increase, followed by a sudden sign change, and then the curve asymptotically approaches zero from below.  This effect can be seen in the excursion to negative $\Delta M$ of the large-$Q$ curves in Figure \ref{fig:deltam}.  The mass scale at which the critical point is reached depends on the magnitude of $|1-r|$; values of $r$ closer to one follow the $\sim$$170\,$MeV limit to larger values of $\hat{M}$.  In the analogous case of radiative mass splittings, $r=M_{\mathrm{pole}}/\hat{M}$, which one would expect to be close to unity unless there are extremely large radiative corrections (which may indicate that unphysical large logarithms are present).

Consider the yellow curves in Figures \ref{fig:pole_masses} and \ref{fig:deltam}, corresponding to $Q=\hat{M}/2$.  Of the scales we consider, this is the best choice of renormalisation scale in terms of minimising unwanted logarithmic corrections, as it interpolates the large mass hierarchy.  As we see in Figure \ref{fig:pole_masses}, this indeed corresponds to a ratio of pole and tree-level masses close to unity at large $\hat{M}$.  In turn, this corresponds to a value of $r\sim 1$ for large $\hat{M}$ in Figure \ref{fig:r_diff}, and thus a suppression of the deviation from the $\sim$$170\,$MeV limit.  The yellow curve on Figure \ref{fig:deltam} illustrates the same behaviour, running closest to the non-iterative result.

This is further verified by considering the other curves in Figure \ref{fig:deltam}.  For $Q$ at the lower end of the scale, at $m_t/2$, the turn-over occurs for relatively small $\hat{M}$.  At the other end of the range for $Q$, where we consider cases with $Q\propto \hat{M}$, once the constant of proportionality becomes sufficiently large a critical point is reached where the turn-over occurs at smaller values of $\hat{M}$, as can be seen between the $Q=\hat{M}$ and $Q=2\hat{M}$ results.  This behaviour is consistent with the idea that this is the result of large logarithms of the form $\log(\hat{m}_{W,Z}/Q)$, $\log(\hat{M}/Q)$ and $\log(\hat{M}_{\mathrm{pole}}/Q)$, which contribute to a large self energy, and are not cancelled in the iterative calculation.  Thus, it is sensible that for the intermediate value of $Q\approx \hat{M}/2$ the iterative mass splitting is in much closer agreement with the non-iterative result, as logarithms from both ends of the hierarchy are better controlled.


\subsection{Gauge choice}

As physical observables, the pole masses and mass splitting should be entirely independent of the gauge choice.  For the non-iterative method we see that this is indeed the case, with consistent results in all three gauges ($\xi = 0,1,3$) \footnote{The running of the $\overline{MS}$ gauge boson masses is not relevant for a one-loop calculation, as the contribution from running is of higher loop order.  We find that the gauge dependence of the running parameters has little effect.  For the calculations discussed in this subsection, we therefore use Feynman gauge for all running, and only change the gauge choice for the self energies.}.  In the iterative method, we see consistency of the mass splitting uncertainty regions between the Landau ($\xi = 0$) and Feynman-'t Hooft ($\xi=1$) gauges, with a slight enlargement in the uncertainty band in the Landau gauge.  The Fried-Yennie gauge ($\xi=3$) shows a significantly larger uncertainty band, although it still overlaps with the other results.

The Fried-Yennie gauge produces the most complicated form for the self energy functions.  The necessary cancellation between the charged and neutral self energies is therefore further compromised by a sensitivity to unwanted logarithmic terms that affect the pole mass.  As seen in Figure \ref{fig:pole_masses}, the pole masses also have the largest uncertainties in this gauge.  In the low-mass range of Figure \ref{fig:deltam}, where $\hat{M}\lesssim m_t$, we see a huge increase in the uncertainty band, driven by solutions with $Q$ at the extreme values of $2m_t$ and $\hat{M}/2$.  The only iterative result in close agreement with the non-iterative mass splitting is the one that interpolates the relatively small mass hierarchy, i.e.\ $Q=m_t/2$.  To a lesser extent, this sensitivity to logarithmic terms can be understood as the reason for the widening of the uncertainty band in the Landau gauge as well.

\subsection{The two-loop mass splitting}\label{sec:two_loop}

With such a large uncertainty in the one-loop mass-splitting, it is of interest to compute the radiative corrections at the next order.  We have computed full two-loop amplitudes for the charged and neutral multiplet components in the Feynman-'t Hooft gauge using the non-iterative method; the details of this calculation are presented elsewhere \cite{McKay2017}.  Without the condition $p^2 = \hat{M}^2$, which is imposed in the non-iterative calculation only, the basis integral reduction fails to produce reliable results, encountering a singularity at $p^2 = \hat{M}^2$ for certain diagrams.  Therefore, we are able to obtain a full two-loop result only with the non-iterative method.  However, in the interests of investigating the behaviour of the iterative calculation at two loops, we have produced a partial two-loop amplitude that can be solved iteratively, based on a combination of diagrams that gives a finite self-energy.  By considering this subgroup of two-loop topologies, we obtained a self energy valid at both $p^2 \neq \hat{M}^2$ and $p^2 = \hat{M}^2$.   The classes of diagrams that we used for this partial two-loop amplitude are summarised in Figure \ref{fig:Feynman_diagrams}; the full set of two-loop diagrams can be found in Figures 1 through 3 of Ref.~\cite{McKay2017}.

\tikzset{
    photon/.style={decorate, decoration={snake,segment length=2mm, amplitude=0.8mm}, draw=black},
    wino/.style={draw=black}
}

\tikzstyle{vertex} = [circle,fill=black,inner sep=0pt,minimum size=3pt]
\tikzstyle{empty} = [circle,fill=none,inner sep=0pt,minimum size=3pt]
\tikzstyle{SMcorrection} = [circle,fill=black,inner sep=0pt,minimum size=15pt]
\tikzstyle{counterterm} = [circle,fill=black,inner sep=0pt,minimum size=10pt]
\newcommand{\CIRCLE}{$\mathbin{\tikz [x=2.28ex,y=2.28ex,line width=.5ex, black] \draw (0,0) -- (1,1) (0,1) -- (1,0);}$}%
\newcommand{\Cross}{$\mathbin{\tikz [x=2.28ex,y=2.28ex,line width=.5ex, gray] \draw (0,0) -- (1,1) (0,1) -- (1,0);}$}%
\newcommand{\CT}{$\mathbin{\tikz [x=1.28ex,y=1.28ex,line width=.5ex, gray] \draw (0,0) -- (1,1) (0,1) -- (1,0);}$}%

\begin{figure*}[t!]
\center{
\setlength\tabcolsep{8pt}
\begin{tabular}{cccccccc}
     \multicolumn{2}{c}{ \addheight{
      \begin{tikzpicture}
\node [vertex] (left) {};
\node [empty,above of = left, node distance = 0.6cm] (left_upper) {};
\node [vertex,right of = left, node distance = 1.8cm] (right) {};
\node [vertex,right of = left_upper, node distance = 0.5cm] (upper_left) {};
\node [vertex,right of = upper_left, node distance = 0.8cm] (upper_right) {};
\node [empty, left of = left, node distance = 0.75cm] (start) {};
\node [empty, right of = right, node distance = 0.75cm] (end) {};
\draw[wino] (start) -- (left);
\draw[wino] (right) -- (end);
\draw[wino] (left) arc (0:90:-0.5cm and 0.6cm);
\draw[photon] (left) arc (0:180:-0.9cm and -0.75cm);
\draw[wino] (upper_left) arc (0:180:-0.4cm and 0.325cm);
\draw[photon] (upper_left) arc (0:180:-0.4cm and -0.325cm);
\draw[wino] (upper_right) arc (270:180:-0.5cm and -0.6cm);
\end{tikzpicture}
      }} &
\multicolumn{4}{c}   {   \addheight{
     \begin{tikzpicture}
\node [vertex] (left) {};
\node [vertex,right of = left, node distance = 1.8cm] (right) {};
\node [empty,above of = left, node distance = 0.75cm] (left_upper) {};
\node [empty,above of = left, node distance = -0.75cm] (left_lower) {};
\node [vertex,right of = left_upper, node distance = 0.9cm] (upper_center) {};
\node [vertex,right of = left_lower, node distance = 0.9cm] (lower_center) {};
\node [empty, left of = left, node distance = 0.75cm] (start) {};
\node [empty, right of = right, node distance = 0.75cm] (end) {};
\draw[wino] (start) -- (left);
\draw[wino] (right) -- (end);
\draw[photon] (left) arc (0:90:-0.9cm and -0.75cm);
\draw[wino] (left) arc (0:90:-0.9cm and 0.75cm);
\draw[photon] (upper_center) arc (270:180:-0.9cm and -0.75cm);
\draw[wino] (lower_center) arc (270:180:-0.9cm and 0.75cm);
\draw[wino] (upper_center) -- (lower_center);
\end{tikzpicture}
      }}&
\multicolumn{2}{c} {     \addheight{
     \begin{tikzpicture}
\node [vertex] (left) {};
\node [vertex,right of = left, node distance = 1.8cm] (right) {};
\node [empty,above of = left, node distance = 0.75cm] (left_upper) {};
\node [empty,above of = left, node distance = -0.75cm] (left_lower) {};
\node [vertex,right of = left_upper, node distance = 0.9cm] (upper_center) {};
\node [vertex,right of = left_lower, node distance = 0.9cm] (lower_center) {};
\node [empty, left of = left, node distance = 0.75cm] (start) {};
\node [empty, right of = right, node distance = 0.75cm] (end) {};
\draw[wino] (start) -- (left);
\draw[wino] (right) -- (end);
\draw[wino] (left) arc (0:90:-0.9cm and -0.75cm);
\draw[photon] (left) arc (0:90:-0.9cm and 0.75cm);
\draw[photon] (upper_center) arc (270:180:-0.9cm and -0.75cm);
\draw[wino] (lower_center) arc (270:180:-0.9cm and 0.75cm);
\draw[photon] (upper_center) -- (lower_center);
\end{tikzpicture}
      }} \\
      \multicolumn{2}{c}   {   \addheight{
      \begin{tikzpicture}
\node [vertex] (left) {};
\node [empty,above of = left, node distance = 0.72cm] (left_upper) {};
\node [vertex,right of = left, node distance = 1.8cm] (right) {};
\node [empty, left of = left, node distance = 0.75cm] (start) {};
\node [empty, right of = right, node distance = 0.75cm] (end) {};
\draw[wino] (start) -- (left);
\draw[wino] (right) -- (end);
\draw[wino] (left) arc (0:180:-0.9cm and 0.75cm);
\draw[photon] (left) arc (0:180:-0.9cm and -0.75cm);
\node [counterterm,right of = left_upper, node distance = 0.9cm] (upper_center) {};
\node [empty,right of = left_upper, node distance = 0.9cm] (upper_center) {\CT};
\end{tikzpicture}
          }}&
\multicolumn{4}{c}   {   \addheight{
      \begin{tikzpicture}
\node [vertex] (left) {};
\node [empty,above of = left, node distance = 0.77cm] (left_upper) {};
\node [counterterm,right of = left, node distance = 1.8cm] (right) {};
\node [empty, left of = left, node distance = 0.75cm] (start) {};
\node [empty, right of = right, node distance = 0.75cm] (end) {};
\draw[wino] (start) -- (left);
\draw[wino] (right) -- (end);
\draw[photon] (left) arc (0:180:-0.9cm and 0.75cm);
\draw[wino] (left) arc (0:180:-0.9cm and -0.75cm);
\node [empty,right of = left, node distance = 1.8cm] (right) {\CT};
\end{tikzpicture}
           }}&
\multicolumn{2}{c} {     \addheight{
      \begin{tikzpicture}
\node [counterterm] (left) {};
\node [empty,above of = left, node distance = 0.77cm] (left_upper) {};
\node [vertex,right of = left, node distance = 1.8cm] (right) {};
\node [empty, left of = left, node distance = 0.75cm] (start) {};
\node [empty, right of = right, node distance = 0.75cm] (end) {};
\draw[wino] (start) -- (left);
\draw[wino] (right) -- (end);
\draw[photon] (left) arc (0:180:-0.9cm and 0.75cm);
\draw[wino] (left) arc (0:180:-0.9cm and -0.75cm);
\node [empty] (left) {\CT};
\end{tikzpicture}
      }} \\

\end{tabular}
}
\caption{Two-loop diagrams contributing to the partial self-energy.  Small circles with crosses indicate counter-term insertions.  Solid lines indicate multiplet fermions ($\chi^{0},\, \chi^{\pm}$) and wiggly lines electroweak vector bosons ($W^{\pm}$, $Z$, $\gamma$).} \label{fig:Feynman_diagrams}
\end{figure*}
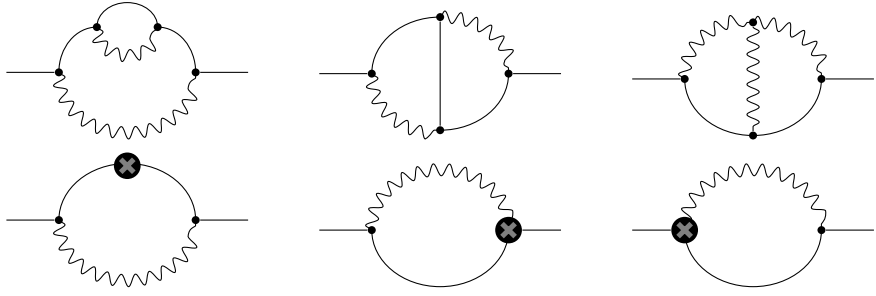

As shown in Ref.\ \cite{Ibe2013}, the two-loop amplitude contains IR divergent terms for $p^2 = \hat{M}^2$.  These divergences cancel with the derivative of the one-loop amplitude when the pole mass is computed using the non-iterative method (Eq.\ \ref{eqn:M_pole_explicit}).  In both the iterative and non-iterative case, we regulate this divergence by using a fictitious -- but small -- mass for the photon, causing the divergences to cancel numerically.  We have verified that the mass splitting is indeed independent of the exact choice for sufficiently small values of the regulator mass.  For the iterative case there is no such cancellation, but the amplitude is IR-safe anyway because $p^2 \neq \hat{M}^2$.  By using a regulator mass even in the iterative calculation, however, we avoid any problem associated with the IR divergence at the first step of the iteration, when $p^2 = \hat{M}^2$.

\begin{figure*}
\centering
\includegraphics[width=0.5\textwidth]{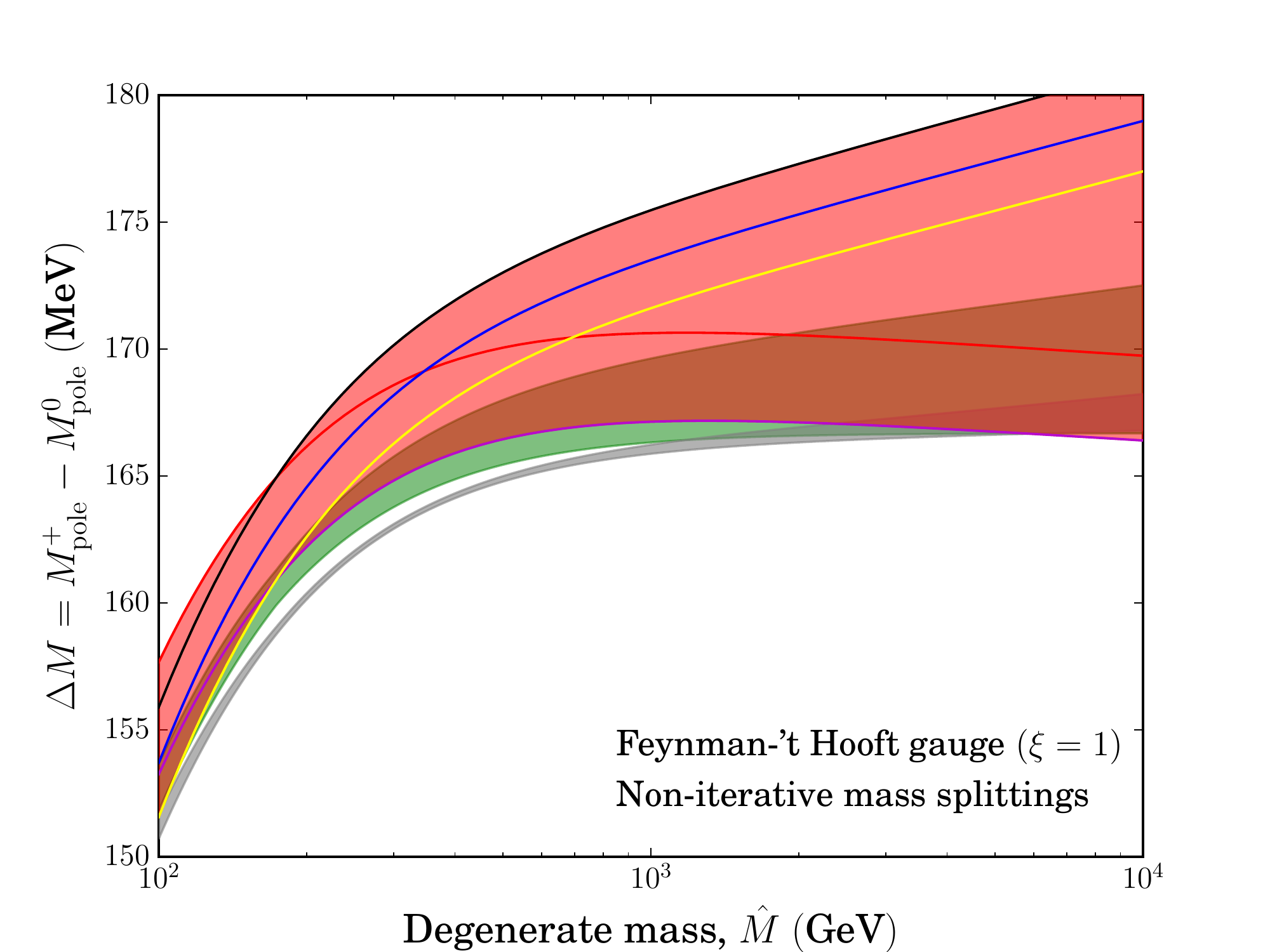}\includegraphics[width=0.5\textwidth]{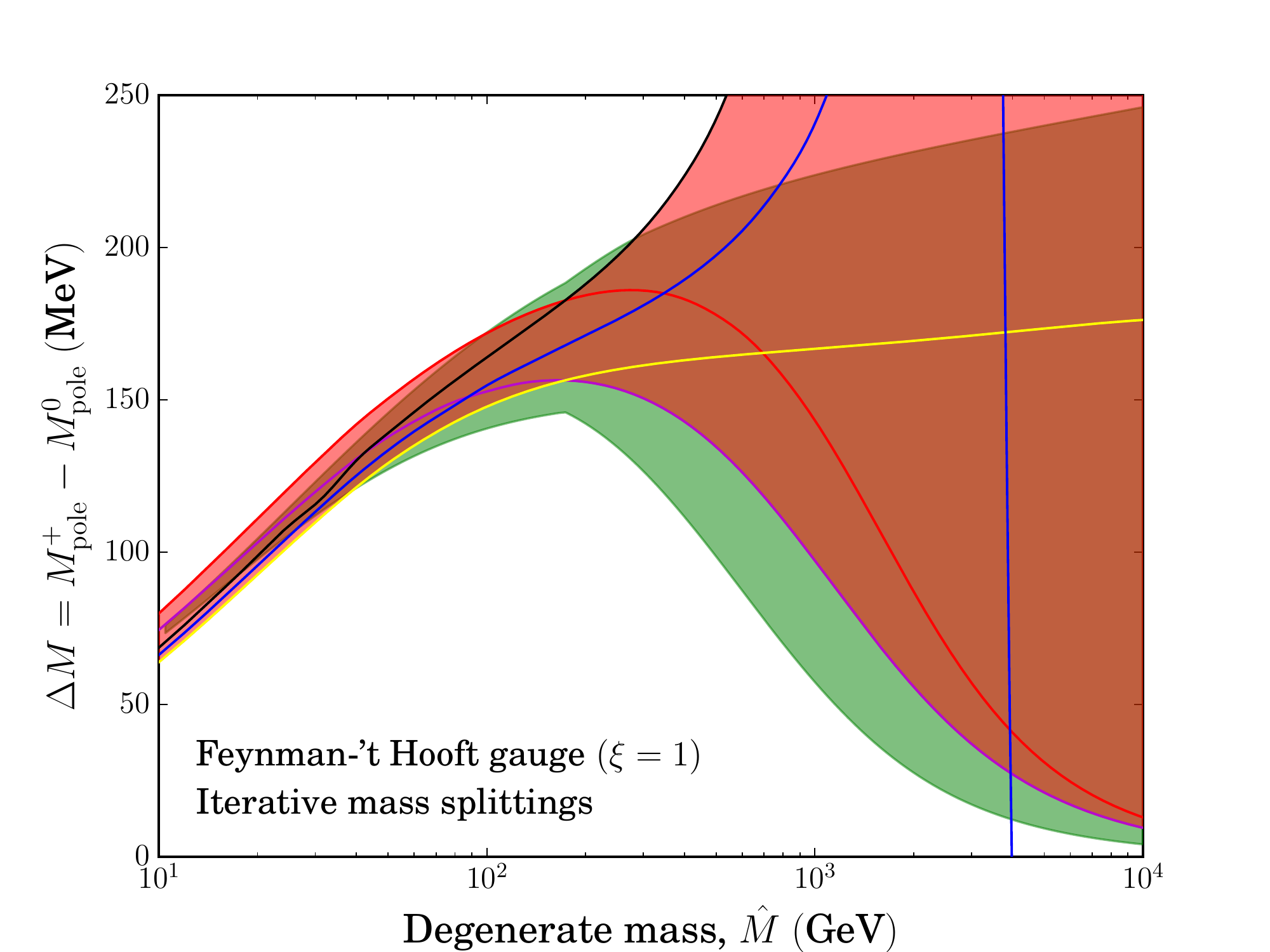}\\
\includegraphics[width=0.7\textwidth]{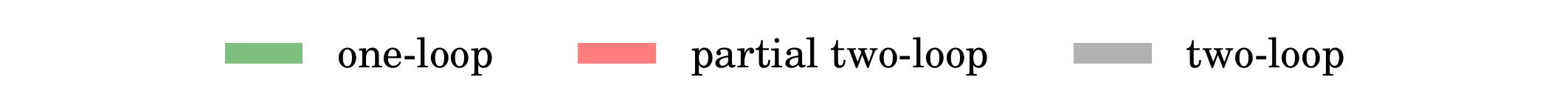}
\includegraphics[width=0.9\textwidth]{legend_1.pdf}
\caption{The splitting $\Delta M\equiv \Mp^+-\Mp^0$ as a function of the degenerate mass $\hat{M}$ at one-loop and two-loop order, for the non-iterative method (\textit{left}) and the iterative method (\textit{right}).  We include partial and full two-loop results for the non-iterative method; only a partial two-loop calculation is possible with the iterative method.  Shaded bands indicate the range of values obtained by varying $Q$ continuously between $\min\{\hat{M}/2,m_t/2\}$ and $\max\{2\hat{M},2m_t\}$.  Solid lines indicate partial two-loop values computed at specific choices of the renormalisation scale $Q$.}\label{fig:deltam_two_loop}
\end{figure*}

In the left panel of Figure \ref{fig:deltam_two_loop}, we begin by comparing the non-iterative mass splitting in the one-loop, partial two-loop and full two-loop calculations. The uncertainty of the full two-loop amplitude due to scale dependence is much smaller than that of the one-loop result.  This confirms findings in the literature, and shows that these cancellations in the non-iterative approach allow the precision of the splitting to improve with the addition of the higher-order contributions, as one would normally expect.  In the limit of large $\hat{M}$, the only remaining $Q$-dependence of the one-loop and full two-loop results comes from renormalisation of the SM input parameters.\footnote{The large $\hat{M}$ limit of the two-loop mass splitting is slightly larger than the result in Ref. \cite{Ibe2013} due to the choice of input parameters.}  Similarly, for solutions at fixed $Q$ and large $\hat{M}$, there is also a very small dependence on $\hat{M}$, seen as a slight decrease in the mass difference with increasing $\hat{M}$; this is due to the influence of the value of $\hat{M}$ on the renormalisation of the SM input parameters.  This can be confirmed by comparison with the left panel of Figure 10 in Ref.~\cite{McKay2017}, where the pink band which is indeed flat for large $\hat{M}$ corresponds to exactly the same result but \textit{without} any threshold corrections applied to the input parameters.

On the other hand, the partial two-loop amplitude shows a relatively large uncertainty, dominated by the results where $Q$ is chosen to be some multiple of $\hat{M}$.  Because even the partial mass splitting is manifestly constant for large $\hat{M}$ and fixed Q (modulo the small and irrelevant impact of running input parameters), we know that there are no terms of the type $\log(\hat{M}/Q)$ left in the result.  We can therefore infer that there are \textit{other} terms proportional to $Q$, of the form $\log(\hat{m}_{Z,W} /Q)$, causing an increase in $\Delta M$ for cases where we have chosen $Q$ to be a multiple of $\hat{M}$.  These terms are clearly cancelled in the full two-loop amplitude, and dominate the uncertainty of our partial two-loop result.  Given the form of these corrections, we can reduce the uncertainty by considering solutions with fixed $Q$ near the electroweak scale.  This gives some control over the effect of the uncancelled scale-dependent terms in the partial mass splitting.  With this constraint, the uncertainty on the non-iterative mass splitting is comparable to the uncertainty of the one-loop result.  This is illustrated by the red and purple lines in the left panel of Figure \ref{fig:deltam_two_loop}, which together bound the uncertainty on the partial result if $Q$ is varied between $m_t/2$ and $2m_t$.

In the right panel of Figure \ref{fig:deltam_two_loop}, we compare the one-loop and partial two-loop mass splittings computed using the iterative procedure.  The behaviour of the partial two-loop mass splitting is reminiscent of the behaviour of the one-loop splitting for $\xi=0,1$ when $Q$ is chosen independently of $\hat{M}$, as seen in Figure \ref{fig:deltam} ($Q=m_t/2,\,2m_t$).  Specifically, the mass splitting converges to zero for large $\hat{M}$, although this occurs at slightly higher masses for the partial two-loop result.  For $Q=\hat{M}/2$ we again see that the iterative mass splitting tracks the $\sim$$170$\,MeV limit reasonably well.  However for $Q>\hat{M}/2$, the partial two-loop splitting extends to very large values (as large as $1$\,GeV for $Q>\hat{M}/2$), much greater than the iterative one-loop mass splitting.  Because we are restricted to a subset of two-loop diagrams that we know in the non-iterative case to result in an \textit{increased} scale-dependence compared to the one-loop result, some increase in the scale-dependence can be expected in the iterative calculation when going from one loop to the partial two-loop result.  Indeed, it is from the solutions for which $Q\propto M$ that we see the large increase in the uncertainty of the explicit mass splitting, so it is not surprising that these solutions lead to a larger uncertainty in the iterative result as well.  For $Q$ chosen independently from $\hat{M}$ ($Q=m_t/2,\,2m_t$), the iterative partial two-loop calculation does show slightly less sensitivity to the renormalisation scale than the one-loop result.  This can be seen by comparing the area bounded by the red and purple curves in the right panels of Figs.\ \ref{fig:deltam} and \ref{fig:deltam_two_loop}.  This suggests that if we were also able to control the uncertainty for solutions with $Q\propto M$ by including the missing diagrams, then the overall uncertainty of the iterative result could be reduced somewhat compared to the one-loop version.

Similarly, the delay of the turnover of the mass splitting to higher multiplet masses, when going from one loop to two, indicates that the two-loop corrections do partially compensate for the large logarithms in $\hat{M}/Q$ responsible for the deviation of the iterative result from the non-iterative one.  However, the asymptotic behaviour for large $\hat{M}$ and $Q=m_t/2,\,2m_t$ remains the same as in the one-loop iterative result, indicating that this compensation is far from complete.  Even with two-loop contributions included, the iterative calculation does not exhibit the cancellation that occurs in the non-iterative case. This suggests that higher-order corrections cannot completely `cure' the scale-dependence of the one-loop iterative calculation, even if they can reduce the effect.

\section{Phenomenological implications}\label{sec:decays}

The precise value of the mass splitting is most relevant in the calculation of the dark matter relic density, and the decay lifetime of the charged component.  We briefly discuss the possible effect of erroneous mass splittings entering into these calculations, in the case that one was to accidentally use an iterative result without being aware of the pitfalls of this method (such a situation may arise if a spectrum generator is used and results passed to other programs without checks in between).

As the typically-assumed $\sim$$170$\,MeV mass splitting is still relatively small compared to the actual pole masses, it has sometimes simply been neglected when calculating the dark matter relic abundance \cite{Cirelli2006}.  In this approximation, the uncertainty on the splitting obviously plays no real role. However, $\Delta M$ is quite important when including the Sommerfeld enhancement \cite{Cohen2013}, as it sets the location of the resonance, so can have a large impact on the resulting relic density at certain values of $\hat{M}$.  This is particularly relevant in the multi-TeV region preferred by the observed relic density.

The calculation of the decay lifetime for the charged component is extremely sensitive to the value of the mass splitting.  To demonstrate the importance of avoiding the iterative method, we compute the effect of the mass splitting uncertainty on the decay lifetime.

The charged component decays as $\chi^+\rightarrow\chi^0+X$, which is dominated by channels where $X$ is either a pion, an electron+neutrino or a muon+neutrino pair.  The decay width for the pion channel in an electroweak $n$-plet is \cite{Cirelli2009,Ibe2013}
\begin{align}
\Gamma_{\pi} = (n^2-1)\frac{G_F^2\Delta M^3 V_{\text{ud}}^2f_{\pi}^3}{4\pi}\sqrt{1-\frac{m_{\pi}^2}{\Delta M^2}}, \label{eqn:pi_width}
\end{align}
where $f_\pi=131$\,MeV, $|V_{\text{ud}}|= 0.97425 \pm 0.00022$ \cite{Patrignani2016} and $m_{\pi}$ is the pion mass.  For $\Delta M\approx 170\,\mathrm{MeV} > m_{\pi}$ this is the dominant decay channel, with a 97.7\% branching ratio \cite{Cirelli2009}.  The other kinematically-available channels are the electron and muon, given by
\begin{align}
\Gamma_{e}=(n^2-1)\frac{G_F^2\Delta M^5}{60\pi^2} \label{eqn:e_width}
\end{align}
and $\Gamma_{\mu}=0.12\Gamma_e$.  The expected lifetime of the charged component is thus $\tau=(\Gamma_e+\Gamma_{\mu}+\Gamma_{\pi})^{-1}$.
\begin{figure}
\centering
\includegraphics[width=0.5\textwidth]{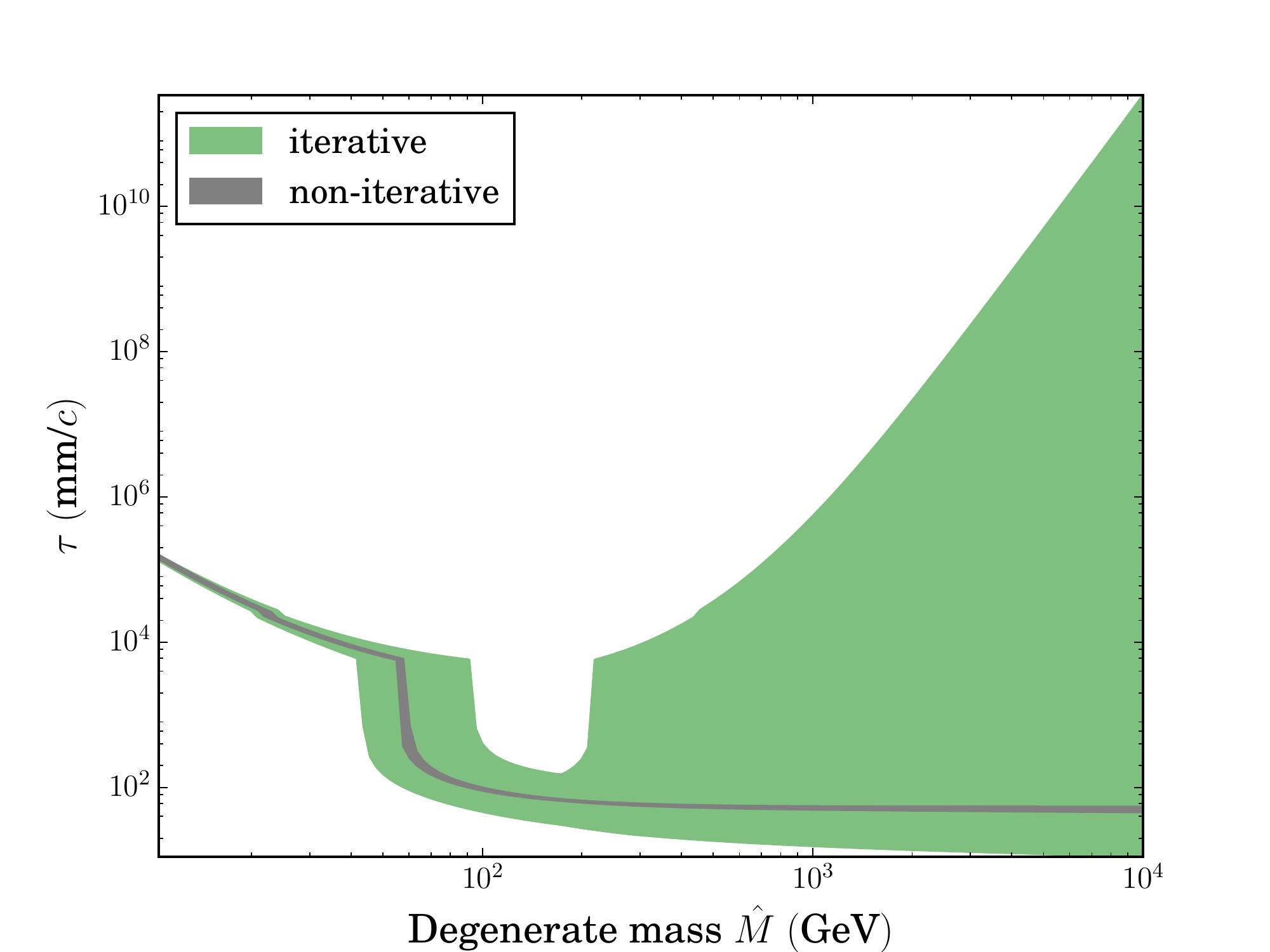}
\caption{
The lifetime of the charged component of an electroweak triplet $\chi^+$, as a function of the degenerate multiplet mass $\hat{M}$, as computed in the Feynman-'t Hooft gauge.  The green and grey regions indicate the range of values obtained by varying $Q$ continuously between $\min\{\hat{M}/2,m_t/2\}$ and $\max\{2\hat{M},2m_t\}$ for the iterative and non-iterative approaches, respectively.  The opening of the pion channel is evident in the large change in the lifetime when $\Delta M = m_{\pi}$, and the opening of the muon channel can be seen in a smaller change at $\Delta M = m_{\mu}$.
}\label{fig:decay}
\end{figure}

In Figure \ref{fig:decay}, we present the decay lifetime of the charged component in units of mm/$c$, as a function of the degenerate mass $\hat{M}$, for both methods of pole mass calculation at one-loop.  The large step in the decay lifetime is where $\Delta M>m_{\pi}$ and the pion channel opens, and the smaller step is due to the muon channel opening.  We can see here that the uncertainty in the one-loop iterative pole mass calculation results in a huge uncertainty in the decay lifetime.  Indeed the width Eq.\ (\ref{eqn:e_width}) has a quintic dependence on the mass splitting.

\section{Conclusion}
\label{sec:conclusion}

In a model where a fermionic multiplet is required to be 100\% of the observed thermal relic abundance of dark matter, the multiplet mass must be of the TeV scale.  Therefore we have $\hat{M}\gg \hat{m}_{W,Z}$ and the calculation of the pole mass involves a large mass hierarchy.  Due to the mathematical form of the non-iterative (or explicit) pole mass, large logarithms of the form $\log(m_X/Q)$, where $m_X\in\{\hat{M},\hat{m}_W,\hat{m}_Z\}$, associated with this hierarchy, cancel out when taking the difference of the charged and neutral components.  Therefore the only renormalisation scale dependence in the mass splittings comes from the input parameters, such as the gauge coupling and running \MSbar masses.

In the iterative method this cancellation is spoiled.  In order to obtain a reasonable estimate of the uncertainty on the resulting mass splitting, such that the iterative and explicit results are consistent, we must vary $Q$ over the entire mass hierarchy.  Despite the fact that both the iterative and explicit pole masses show almost identical variance with respect to the renormalisation scale, only the difference of the iterative masses suffers from a similarly large uncertainty.  This is the result of cancellations not occurring in the iterative case, due to the nature of the procedure.  Any renormalisation scale dependence in the explicit mass splitting is cancelled out perfectly.

The iterative method of calculating a pole mass is a natural choice for a computer program.  As demonstrated in Section \ref{sec:pole_masses}, it uses the least approximations and is straightforward to implement at any loop order.  As seen in Figure \ref{fig:pole_masses}, the choice of either iterative or non-iterative calculation is typically not consequential for pole masses themselves, so it is understandable that publicly-available spectrum generators have made different choices over which approach to use.  Of particular relevance here, \sarahs/\spheno \cite{Porod:2011nf,Staub:2012pb,Staub2014} spectrum generators use an iterative procedure for all pole masses, with no alternative option, whereas \flexiblesusy \cite{Athron2015} enables the user to select either \textit{high precision} (iterative), or \textit{medium/low precision} (non-iterative).  Although the \flexiblesusy names for these options imply that the iterative method is more precise, we can see from Figure \ref{fig:deltam} that this is certainly not always the case for differences between pole masses.

We have reproduced the large uncertainty in the mass splitting using pole masses computed with both \flexiblesusy and \spheno with the iterative method.  We have also reproduced the non-iterative result using \flexiblesusys's \textit{low precision} mode.

A pertinent question is if one must consider the uncertainty arising from the iterative result when using calculations of electroweak mass splittings for doing phenomenology.  On the basis of our investigations in this paper, we argue that this is not necessary.  Due to fortunate cancellations, the explicit method is able to predict the mass difference while being free from logarithmic terms containing explicit scale dependences.  From a physical point of view, with this method we are able to minimise the sensitivity of the final result to non-physical renormalisation-scale effects.  Furthermore, a finite mass splitting is predicted by a classical effect -- the Coulomb energy.  Ref.\ \cite{Cirelli2006} shows agreement between the classical prediction and the value derived from the self energies in Eq.\ (\ref{eqn:simple_mass_splitting}).  Relying on a classical argument alone is of course not sufficient to safely disregard the large uncertainty of the iterative result.  By understanding however that the origin of this uncertainty lies in scale dependence, and that this can be safely removed by performing the explicit calculation, we can safely conclude that the explicit result is indeed accurate to within its own error margin, and should therefore be adopted as such for phenomenological analyses.

\begin{acknowledgements}
We would like to thank Masahiro Ibe, Shigeki Matsumoto, Ryosuke Sato and Dominik St\"ockinger for useful discussions, Vladyslav Shtabovenko and Stephen Martin for helpful correspondence regarding technical aspects of two-loop calculations and the use of \feyncalc and \tsils, respectively.  We thank Florian Staub for helpful correspondence.  JM is supported by the Imperial College London President's PhD Scholarship, PS by STFC (ST/K00414X/1, ST/P000762/1) and PA by the Australian Research Council (CE110001004, FT160100274).
\end{acknowledgements}

\appendix

\section{One-loop self energies}\label{sec:self_energies}

Here we present the one-loop self energies for the charged and neutral components of the multiplet in a general gauge, parameterised by the gauge parameter $\xi$.  We first define the basis integrals required and then write down the full one-loop self energies and counter-terms.

The self energies are written in terms of Passarino-Veltman (PV) \cite{tHooft1979,Passarino1979} functions.
In this investigation we need only two of these integrals, which are defined as
\begin{align}
&A_0(m) = 16\pi^2Q^{4-d}\int{d^dq\over i\,(2\pi)^d}{1\over
q^2+m^2+i\varepsilon}\\
&B_0(p, m_1, m_2) =\nonumber\\
&16\pi^2Q^{4-d}\int{d^dq\over i\,(2\pi)^d}
{1\over\left[q^2+m^2_1+i\varepsilon\right]\left[
(q-p)^2+m_2^2+i\varepsilon\right]},
\label{B0 def}
\end{align}
where we use $d=4-2\epsilon$ and we will hereafter let $B_0(p, m_1, m_2) = B_0(m_1, m_2)$.
These complex functions can be expressed analytically. $B_0$ is given by Pierce, Bagger and Matchev \cite{Pierce1997} as
\begin{align}
B_0(p,m_1,m_2)=\Delta-\log\left(\frac{p^2}{Q^2}\right)-f_B(x_+)-f_B(x_-),\label{eqn:B0_analytic}
\end{align}
where $\Delta=2/(4-d)-\gamma+\log(4\pi)$,
\begin{align}
x_{\pm}=&\frac{s\pm\sqrt{s^2-4p^2(m_1^2-i\epsilon)}}{2p^2}, \\ f_B(x)=&\log(1-x)-x\log(1-x^{-1})-1
\end{align}
and $s=p^2-m_2^2+m_1^2$.  The $A_0$ function can be integrated to give
\begin{align}
A_0(m)=m^2\left(\log\left(\frac{m^2}{Q^2}\right)-1+\Delta\right).
\end{align}

The self energy of the charged component $\mychi^+$ is given by (henceforth omitting the $1/(16\pi^2)$ pre-factor from all self energies)

\begin{align}
\begin{split}
\Sigma^+(p^2) = \, & C^+_{A_\chi}\,A(\hat{M})  + C^+_{A_W}\,A(\hat{m}_W)   + C^+_{A_Z}\,A(\hat{m}_Z)        \\
& + C^+_{A_{W\xi}}\,A(\xi\hat{m}_W)   + C^+_{A_{Z\xi}}\,A(\xi\hat{m}_Z)         \\
&   +  C^+_{B_{\chi\gamma}}\,B(\hat{M},0)     + C^+_{B_{\chi W}}\,B(\hat{M},\hat{m}_W)\\
&  + C^+_{B_{\chi W\xi}}\,B(\hat{M},\xi\hat{m}_W)      + C^+_{B_{\chi Z\xi}}\,B(\hat{M},\xi\hat{m}_Z)\\
&              + C^+_{B_{\chi Z}}\,B(\hat{M},\hat{m}_Z)    + C^+_0,
\end{split}
\end{align}
with coefficients
\small{
\begin{eqnarray}
&C^+_{A_{\chi}}  & = \frac{g^2}{2p^2} \left[ (\xi^2-1)\cos(2\theta)-\xi^2-3\right] \slashed{p} \\
&C^+_{A_W}  & =  \frac{g^2}{2p^2} \left( 2\hat{m}_W^2+\hat{M}^2-p^2   \right) \slashed{p}   \\
&C^+_{A_Z}  & =  \frac{g^2\cos^2(\theta)}{2p^2} \left(  2\hat{m}_Z^2+\hat{M}^2-p^2   \right) \slashed{p}   \\
&C^+_{A_{W\xi}}  & =  \frac{g^2}{2p^2} \left( p^2-\hat{M}^2   \right) \slashed{p}   \\
&C^+_{A_{Z\xi}}  & =  \frac{g^2\cos^2(\theta)}{2p^2} \left(   p^2-\hat{M}^2    \right) \slashed{p}   \\
&C^+_{B_{\chi\gamma}}  & =  -\frac{g^2\sin^2(\theta)\xi^2}{p^2} \left(  p^2+\hat{M}^2    \right) \slashed{p}
- g^2\sin^2(\theta) \left( \xi^2+3  \right) \hat{M}  \\
&C^+_{B_{\chi W}}  & = -\frac{g^2}{2\hat{m}_W^2p^2}\left[ p^2\left(p^2-2\hat{M}^2+\hat{m}_W^2\right)+\hat{M}^4+\hat{M}^2\hat{m}_W^2-\hat{m}_W^4\right] \slashed{p} \nonumber\\
&& \phantom{=} - 3g^2\hat{M}  \\
&C^+_{B_{\chi Z}}  & =   -\frac{g^2\cos^2(\theta)}{2\hat{m}_Z^2p^2}\left[ p^2\left(p^2-2\hat{M}^2+\hat{m}_Z^2\right)+\hat{M}^4+\hat{M}^2\hat{m}_Z^2-\hat{m}_Z^4\right] \slashed{p}\nonumber\\
&& \phantom{=} - 3g^2\cos^2(\theta)\hat{M}  \\
&C^+_{B_{\chi W\xi}}  & = -\frac{g^2}{2\hat{m}_W^2p^2} \left[ p^2 \left(2\hat{M}^2-p^2+\hat{m}_W^2\xi^2\right)-\hat{M}^4+\hat{M}^2\hat{m}_W^2\xi^2\right]\slashed{p}\nonumber\\
&&\phantom{=} -g^2\hat{M}\xi^2  \\
&C^+_{B_{\chi Z\xi}}  & =   -\frac{g^2\cos^2(\theta)}{2\hat{m}_Z^2p^2} \left[ p^2 \left(2\hat{M}^2-p^2+\hat{m}_Z^2\xi^2\right)-\hat{M}^4+\hat{M}^2\hat{m}_Z^2\xi^2\right]\slashed{p}\nonumber\\
&&\phantom{=} -g^2\cos^2(\theta)\hat{M}\xi^2  \\
&C^+_0   & =  \left\{\frac{g^2}{2}\left[ \left(\xi^2-1\right)\cos(2\theta)-\xi^2-3\right]+\delta_Z\right\}\slashed{p} \nonumber\\
&&\phantom{=}+ \left(4g^2 + \delta_M\right)\hat{M}.
\end{eqnarray}
}
The self energy of the neutral component $\chi^0$ is
\begin{align}
\begin{split}
\Sigma^0(p^2) =& \, C^0_{A_\chi}\,A(\hat{M})  + C^0_{A_W}\,A(\hat{m}_W)  + C^0_{A_{W\xi}}\,A(\xi\hat{m}_W) + C^0_{B_{\chi W}}\,B(\hat{M},\hat{m}_W)  \\ &+ C^0_{B_{\chi W\xi}}\,B(\hat{M},\xi\hat{m}_W) + C^0_0
\end{split}
\end{align}
 with coefficients
 \begin{eqnarray}
&C^0_{A_\chi} &= -\frac{2g^2}{p^2} \slashed{p} \\
&C^0_{A_W} &= \frac{g^2}{\hat{m}_W^2 p^2} \left( \hat{M}^2-p^2+2\hat{m}_W^2\right) \slashed{p}\\
&C^0_{A_{W\xi}} &= - \frac{g^2}{\hat{m}_W^2p^2}\left(\hat{M}^2-p^2\right)\slashed{p} \\
&C^0_{B_{\chi W}} &= -\frac{g^2}{\hat{m}_W^2p^2} \left( -2\hat{M}^2p^2+\hat{m}_W^2p^2+p^4+\hat{M}^4+\hat{M}^2\hat{m}_W^2-2\hat{m}_W^2\right)\slashed{p} \nonumber\\
&&\phantom{=}- 6g^2\hat{M}\\
&C^0_{B_{\chi W\xi}} &=  \frac{g^2}{\hat{m}_W^2p^2} \left( -2\hat{M}^2p^2-\hat{m}_W^2\xi^2p^2+p^4+\hat{M}^4-\hat{M}^2\hat{m}_W^2\xi^2\right)\slashed{p}  \nonumber\\
&&\phantom{=}- 2g^2\xi^2 \hat{M} \\
&C^0_0 &= \left(-2g^2 + \delta_Z\right) \slashed{p}  + \left(4g^2+\delta_M\right)\hat{M}.
\end{eqnarray}
The separation of the self energy into the form $\Sigma(p^2)=\Sigma_K(p^2)\slashed{p}+\Sigma_M(p^2)$ is manifest in the form of the coefficients presented above.  The counter-terms $\delta_Z$ and $\delta_M$ required to cancel divergences arising from $B_0$ and $A_0$ are
\begin{eqnarray}
\delta_Z&=&4g^2\xi^2\Delta,\\
\delta_M&=&-4g^2(\xi^2+3)\Delta.
\end{eqnarray}


\bibliography{../library}{}

\providecommand{\href}[2]{#2}\begingroup\raggedright\begin{thebibliography}{10}

\bibitem{Hisano2007}
J.~{Hisano}, S.~{Matsumoto}, M.~{Nagai}, O.~{Saito}, and M.~{Senami}, {\it
  {Non-perturbative effect on thermal relic abundance of dark matter}},  {\em
  Physics Letters B} {\bf 646} (2007) 34--38,
  [\href{http://arxiv.org/abs/hep-ph/0610249}{{\tt hep-ph/0610249}}].

\bibitem{Hryczuk2011}
A.~{Hryczuk}, R.~{Iengo}, and P.~{Ullio}, {\it {Relic densities including
  Sommerfeld enhancements in the MSSM}},  {\em Journal of High Energy Physics}
  {\bf 3} (2011) 69, [\href{http://arxiv.org/abs/1010.2172}{{\tt
  arXiv:1010.2172}}].

\bibitem{Cheng1999}
H.-C. {Cheng}, B.~A. {Dobrescu}, and K.~T. {Matchev}, {\it {Generic and chiral
  extensions of the supersymmetric standard model}},  {\em Nuclear Physics B}
  {\bf 543} (1999) 47--72, [\href{http://arxiv.org/abs/hep-ph/9811316}{{\tt
  hep-ph/9811316}}].

\bibitem{Feng1999}
J.~L. {Feng}, T.~{Moroi}, L.~{Randall}, M.~{Strassler}, and S.~{Su}, {\it
  {Discovering Supersymmetry at the Tevatron in W-ino Lightest Supersymmetric
  Particle Scenarios}},  {\em Physical Review Letters} {\bf 83} (1999)
  1731--1734, [\href{http://arxiv.org/abs/hep-ph/9904250}{{\tt
  hep-ph/9904250}}].

\bibitem{Ibe2013}
M.~{Ibe}, S.~{Matsumoto}, and R.~{Sato}, {\it {Mass splitting between charged
  and neutral winos at two-loop level}},  {\em Physics Letters B} {\bf 721}
  (2013) 252--260, [\href{http://arxiv.org/abs/1212.5989}{{\tt
  arXiv:1212.5989}}].

\bibitem{Cirelli2006}
M.~{Cirelli}, N.~{Fornengo}, and A.~{Strumia}, {\it {Minimal dark matter}},
  {\em Nuclear Physics B} {\bf 753} (2006) 178--194,
  [\href{http://arxiv.org/abs/hep-ph/0512090}{{\tt hep-ph/0512090}}].

\bibitem{Cirelli2009}
M.~{Cirelli} and A.~{Strumia}, {\it {Minimal dark matter: model and results}},
  {\em New Journal of Physics} {\bf 11} (2009) 105005,
  [\href{http://arxiv.org/abs/0903.3381}{{\tt arXiv:0903.3381}}].

\bibitem{Cai2015}
C.~{Cai}, Z.-M. {Huang}, Z.~{Kang}, Z.-H. {Yu}, and H.-H. {Zhang}, {\it
  {Perturbativity limits for scalar minimal dark matter with Yukawa
  interactions: Septuplet}},  {\em \prd} {\bf 92} (2015) 115004,
  [\href{http://arxiv.org/abs/1510.01559}{{\tt arXiv:1510.01559}}].

\bibitem{Cirelli2007}
M.~{Cirelli}, A.~{Strumia}, and M.~{Tamburini}, {\it {Cosmology and
  astrophysics of minimal dark matter}},  {\em Nuclear Physics B} {\bf 787}
  (2007) 152--175, [\href{http://arxiv.org/abs/0706.4071}{{\tt
  arXiv:0706.4071}}].

\bibitem{Chen2012a}
C.-S. {Chen} and Y.~{Tang}, {\it {Vacuum stability, neutrinos, and dark
  matter}},  {\em Journal of High Energy Physics} {\bf 4} (2012) 19,
  [\href{http://arxiv.org/abs/1202.5717}{{\tt arXiv:1202.5717}}].

\bibitem{DiLuzio2015}
L.~{Di Luzio}, R.~{Gr{\"o}ber}, J.~F. {Kamenik}, and M.~{Nardecchia}, {\it
  {Accidental matter at the LHC}},  {\em Journal of High Energy Physics} {\bf
  7} (2015) 74, [\href{http://arxiv.org/abs/1504.00359}{{\tt
  arXiv:1504.00359}}].

\bibitem{Belyaev2017}
A.~Belyaev, G.~Cacciapaglia, J.~McKay, D.~Marin, and A.~R. Zerwekh, {\it
  {Minimal Spin-one Isotriplet Dark Matter}},  {\em In preparation} (2017).

\bibitem{Ostdiek2015}
B.~Ostdiek, {\it {Constraining the minimal dark matter fiveplet with LHC
  searches}},  {\em Physical Review D} {\bf 92} (2015) 055008.

\bibitem{Porod:2011nf}
W.~Porod and F.~Staub, {\it {SPheno 3.1: Extensions including flavour,
  CP-phases and models beyond the MSSM}},  {\em Comput. Phys. Commun.} {\bf
  183} (2012) 2458--2469, [\href{http://arxiv.org/abs/1104.1573}{{\tt
  arXiv:1104.1573}}].

\bibitem{Staub:2012pb}
F.~Staub, {\it {SARAH 3.2: Dirac Gauginos, UFO output, and more}},  {\em
  Comput. Phys. Commun.} {\bf 184} (2013) 1792--1809,
  [\href{http://arxiv.org/abs/1207.0906}{{\tt arXiv:1207.0906}}].

\bibitem{Staub2014}
F.~{Staub}, {\it {SARAH 4: A tool for (not only SUSY) model builders}},  {\em
  Computer Physics Communications} {\bf 185} (2014) 1773--1790,
  [\href{http://arxiv.org/abs/1309.7223}{{\tt arXiv:1309.7223}}].

\bibitem{Athron2015}
P.~{Athron}, J.-h. {Park}, D.~{St{\"o}ckinger}, and A.~{Voigt}, {\it
  {FlexibleSUSY-A spectrum generator generator for supersymmetric models}},
  {\em Computer Physics Communications} {\bf 190} (2015) 139--172,
  [\href{http://arxiv.org/abs/1406.2319}{{\tt arXiv:1406.2319}}].

\bibitem{DelNobile2010}
E.~{Del Nobile}, R.~Franceschini, D.~Pappadopulo, and A.~Strumia, {\it {Minimal
  Matter at the Large Hadron Collider}},  {\em Nuclear Physics B} {\bf 826}
  (2010) 217--234, [\href{http://arxiv.org/abs/0908.1567}{{\tt
  arXiv:0908.1567}}].

\bibitem{Yamada2010}
Y.~{Yamada}, {\it {Electroweak two-loop contribution to the mass splitting
  within a new heavy SU(2$_{}$ fermion multiplet}},  {\em Physics Letters B}
  {\bf 682} (2010) 435--440, [\href{http://arxiv.org/abs/0906.5207}{{\tt
  arXiv:0906.5207}}].

\bibitem{McKay2017}
J.~McKay and P.~Scott, {\it Two-loop mass splittings in electroweak multiplets:
  Winos and minimal dark matter},  {\em Phys. Rev. D} {\bf 97} (2018) 055049,
  [\href{http://arxiv.org/abs/1712.00968}{{\tt arXiv:1712.00968}}].

\bibitem{Mertig1991}
R.~Mertig, M.~B{\"{o}}hm, and A.~Denner, {\it {Feyn Calc - Computer-algebraic
  calculation of Feynman amplitudes}},  {\em Computer Physics Communications}
  {\bf 64} (1991) 345--359.

\bibitem{Shtabovenko2016}
V.~{Shtabovenko}, R.~{Mertig}, and F.~{Orellana}, {\it {New developments in
  FeynCalc 9.0}},  {\em Computer Physics Communications} {\bf 207} (2016)
  432--444, [\href{http://arxiv.org/abs/1601.01167}{{\tt arXiv:1601.01167}}].

\bibitem{Hahn2001}
T.~{Hahn}, {\it {Generating Feynman diagrams and amplitudes with FeynArts 3}},
  {\em Computer Physics Communications} {\bf 140} (2001) 418--431,
  [\href{http://arxiv.org/abs/hep-ph/0012260}{{\tt hep-ph/0012260}}].

\bibitem{Smirnov2015}
A.~V. {Smirnov}, {\it {FIRE5: A C++ implementation of Feynman Integral
  REduction}},  {\em Computer Physics Communications} {\bf 189} (2015)
  182--191, [\href{http://arxiv.org/abs/1408.2372}{{\tt arXiv:1408.2372}}].

\bibitem{SHTABOVENKO201748}
V.~{Shtabovenko}, {\it {FeynHelpers: Connecting FeynCalc to FIRE and
  Package-X}},  {\em Computer Physics Communications} {\bf 218} (2017) 48--65,
  [\href{http://arxiv.org/abs/1611.06793}{{\tt arXiv:1611.06793}}].

\bibitem{Mertig1998}
R.~{Mertig} and R.~{Scharf}, {\it {TARCER - A mathematica program for the
  reduction of two-loop propagator integrals}},  {\em Computer Physics
  Communications} {\bf 111} (1998) 265--273,
  [\href{http://arxiv.org/abs/hep-ph/9801383}{{\tt hep-ph/9801383}}].

\bibitem{Martin2006}
S.~P. {Martin} and D.~G. {Robertson}, {\it {TSIL: a program for the calculation
  of two-loop self-energy integrals}},  {\em Computer Physics Communications}
  {\bf 174} (2006) 133--151, [\href{http://arxiv.org/abs/hep-ph/0501132}{{\tt
  hep-ph/0501132}}].

\bibitem{Pierce1997}
D.~M. {Pierce}, J.~A. {Bagger}, K.~T. {Matchev}, and R.-J. {Zhang}, {\it
  {Precision corrections in the minimal supersymmetric standard model}},  {\em
  Nuclear Physics B} {\bf 491} (1997) 3--67,
  [\href{http://arxiv.org/abs/hep-ph/9606211}{{\tt hep-ph/9606211}}].

\bibitem{Staub:2009bi}
F.~Staub, {\it {From Superpotential to Model Files for FeynArts and
  CalcHep/CompHep}},  {\em Comput.Phys.Commun.} {\bf 181} (2010) 1077--1086,
  [\href{http://arxiv.org/abs/0909.2863}{{\tt arXiv:0909.2863}}].

\bibitem{Staub:2010jh}
F.~Staub, {\it {Automatic Calculation of supersymmetric Renormalization Group
  Equations and Self Energies}},  {\em Comput.Phys.Commun.} {\bf 182} (2011)
  808--833, [\href{http://arxiv.org/abs/1002.0840}{{\tt arXiv:1002.0840}}].

\bibitem{Allanach2002}
B.~C. {Allanach}, {\it {SOFTSUSY: A program for calculating supersymmetric
  spectra}},  {\em Computer Physics Communications} {\bf 143} (2002) 305--331,
  [\href{http://arxiv.org/abs/hep-ph/0104145}{{\tt hep-ph/0104145}}].

\bibitem{Allanach:2013kza}
B.~Allanach, P.~Athron, L.~C. Tunstall, A.~Voigt, and A.~Williams, {\it
  {Next-to-Minimal SOFTSUSY}},  {\em Comput.Phys.Commun.} {\bf 185} (2014)
  2322--2339, [\href{http://arxiv.org/abs/1311.7659}{{\tt arXiv:1311.7659}}].

\bibitem{Romao2006}
J.~C. Romao, {\it {Modern Techniques for One-Loop Calculations}},  2006.

\bibitem{tHooft1979}
G.~{'t Hooft} and M.~Veltman, {\it {Scalar One Loop Integrals}},  {\em Nuclear
  Physics B} {\bf 153} (1979) 365--401.

\bibitem{Passarino1979}
G.~Passarino and M.~Veltman, {\it {One-loop corrections for $e^+e^-$
  annihilation into $\mu^+\mu^-$ in the Weinberg model}},  {\em Nuclear Physics
  B} {\bf 160} (1979) 151 -- 207.

\bibitem{Cohen2013}
T.~Cohen, M.~Lisanti, A.~Pierce, and T.~R. Slatyer, {\it {Wino dark matter
  under siege}},  {\em Journal of Cosmology and Astroparticle Physics} {\bf 10}
  (2013) 61.

\bibitem{Patrignani2016}
{Particle Data Group}, {\it {Review of Particle Physics}},  {\em Chinese
  Physics C} {\bf 40} (2016) 100001.

\end{thebibliography}\endgroup
\bibliographystyle{bibstyle}

\end{document}